\documentclass[12pt]{article}
\usepackage[utf8]{inputenc}

\pdfoutput=1

\usepackage[margin=1.25in]{geometry}

\usepackage{tikz}
\usetikzlibrary{calc,arrows.meta}
\usepackage{subcaption}
\usepackage[round]{natbib}
\usepackage{amssymb,amsmath,amsthm}
\usepackage{mathtools}
\usepackage{algorithm}
\usepackage{algpseudocode}
\usepackage{nicefrac}
\usepackage{setspace}
\usepackage{soul}
\usepackage[nottoc]{tocbibind}

\allowdisplaybreaks

\usepackage[
  bookmarks=true,
  bookmarksnumbered=true,
  bookmarksopen=true,
  pdfborder={0 0 0},
  breaklinks=true,
  colorlinks=true,
  linkcolor=black,
  citecolor=black,
  filecolor=black,
  urlcolor=blue,
]{hyperref}
\usepackage[capitalise]{cleveref}
\Crefname{figure}{Figure}{Figures}

\theoremstyle{plain}
\newtheorem{theorem}{Theorem}

\newtheorem{proposition}{Proposition}
\newtheorem{lemma}{Lemma}
\theoremstyle{definition}
\newtheorem{definition}{Definition}
\newtheorem{axiom}{Axiom}
\Crefname{axiom}{Axiom}{Axioms}

\newtheorem{example}{Example}

\newcommand{\eqdef}{\triangleq}

\DeclareMathOperator{\rad}{rad}
\DeclareMathOperator{\Pareto}{Pareto}
\newcommand{\compose}{\otimes}

\hypersetup{
  pdftitle       = {Quantifying Inefficiency},
  pdfauthor      = {Yannai A. Gonczarowski <yannai@gonch.name> and Ella Segev <ella.segev@mail.huji.ac.il>}
}

\title{Quantifying Inefficiency\thanks{
Gonczarowski: Department of Economics and Department of Computer Science, Harvard University (\emph{e-mail}: \mbox{\href{mailto:yannai@gonch.name}{yannai@gonch.name}}); Segev: The Hebrew University Business School and the Federmann Center for the Study of Rationality, The Hebrew University of Jerusalem (\emph{e-mail}: \mbox{\href{mailto:ella.segev@mail.huji.ac.il}{ella.segev@mail.huji.ac.il}}).
We thank Cassidy Shubatt for excellent research assistance, and thank Gabriel Carroll, Francesco Fabbri, Loren Fryxell, Jerry Green, Sergiu Hart, Scott Kominers, Eric Maskin, Debraj Ray, Tim Roughgarden, Assaf Romm, Itai Sher, Ran Shorrer, Tomasz Strzalecki, Omer Tamuz, Rakesh Vohra, and Frank Yang for insightful comments and discussions. We thank participants at the Georgetown Microeconomic Theory Seminar, the Normative Economics and Economic Policy Webinar, the University of Notre Dame Economic Theory Mini-Conference, the Harvard--MIT Economic Theory Seminar, the Caltech Social Sciences Seminar, the Hebrew University Economic Theory Seminar, the Warwick Economic Theory Workshop, the A16z Crypto research seminar series, the Harvard Theory of Computation Seminar, the Northeastern Computer~Science Theory Seminar, the Penn State Microeconomic Theory Seminar, and the Stony Brook Workshop on New Perspectives on Algorithmic Game Theory, for helpful comments. We gratefully acknowledge research support by the following sources. Gonczarowski: National Science Foundation (NSF-BSF grant No.\ 2343922), Harvard FAS Inequality in America Initiative, and Harvard FAS Dean's Competitive Fund for Promising Scholarship. Segev: Israel Science Foundation (ISF grant No.\ 301/23).}}
\author{Yannai A.\ Gonczarowski \and Ella Segev}

\date{August 20, 2026}

\begin{document}

\pagenumbering{roman}

\begin{titlepage}
    
\maketitle

\begin{abstract}
We axiomatically define a \emph{cardinal} social inefficiency function, which, given alternatives and individuals' vNM preferences, assigns a number---the \emph{social inefficiency}---to each alternative. These numbers induce not only an ordinal comparison but also a cardinal comparison between alternatives that is uniquely defined by our axioms despite no exogenously given interpersonal comparison, outside option, or disagreement point. We interpret these numbers as per-capita losses in endogenously normalized utility. We apply our social inefficiency function to a setting where interpersonal comparison is hard to justify---object allocation without money---leveraging techniques from computer science to prove an approximate-efficiency result for the Random Serial Dictatorship mechanism.
\end{abstract}

\thispagestyle{empty}

\clearpage

\thispagestyle{empty}

\tableofcontents

\end{titlepage}

\pagenumbering{arabic}

\section{Introduction}

The question of how to aggregate the preferences of individual members of a society over a set of possible alternatives into a social order is at the heart of social choice theory, dating back to \cite{arrow1951social} and \cite{harsanyi1955cardinal}.
Such ordinal aggregation allows us to compare any two alternatives to determine which of the two is socially better, yet does not generally allow us to quantify the \emph{magnitude} by which one alternative is socially better than another. 
For example, such ordinal aggregation allows us to identify whether a given alternative is socially optimal; however, in case a given alternative is not socially optimal, such ordinal aggregation does not generally allow us to quantify \emph{how far} this alternative is from being optimal---i.e., \emph{how socially inefficient} this alternative is. In this paper, we fill this lacuna.\looseness=-1

Consider an individual's ordinal preferences (over lotteries over alternatives). Under the standard von Neumann--Morgenstern (vNM) axioms, these preferences have an expected-utility representation, that is, utility values for each pure alternative, such that a lottery over alternatives is preferred over another if and only if its expected utility is higher. However, these utilities are only unique up to a monotone affine transformation: Multiplying all the utilities of any specific individual by a positive constant, or shifting them by a constant, does not change the preferences that they represent.\footnote{While some papers use the term \emph{cardinal preferences} for such preferences (whose utility representation is unique up to a monotone affine transformation), in this paper we call these \emph{ordinal preferences} to emphasize that any utility representation of them captures no more than an order over lotteries. We reserve the term \emph{cardinal} for measures that capture richer information beyond only orders.\looseness=-1} This underscores the problematic nature of interpersonal comparability of utilities, which is precluded by the mainstream view within social choice theory \citep{arrow1951social}. Beyond the challenges that the preclusion of interpersonal comparability poses for the ordinal aggregation of preferences, it is even less clear how one might go about defining a notion of \emph{how socially inefficient} some alternative is while depending only on individuals' ordinal preferences and not on any arbitrary choice of affine transformation/normalization of utilities.

In this paper, despite the aforementioned challenges, we axiomatically define not an ordinal but a \emph{cardinal} measure of the social (in)efficiency of any given alternative. This measure provides a unique cardinal comparison between alternatives---even across different settings, e.g., allowing one to numerically compare the social (in)efficiency of the different equilibria that arise in a game for different preference profiles. And yet, this measure is based solely on the (ordinal) von Neumann--Morgenstern (vNM) preferences of the individuals. 

One of the early major contributions of theoretical computer science to economic theory is in its abundant use of 
(and techniques for deriving) \emph{inefficiency-bounding theorems}.\footnote{These are many times referred to as \emph{approximation theorems} within computer science.} Such theorems, rather than ascertaining the efficiency of some equilibrium, mechanism, or strategy, ascertain that it is guaranteed to not be far from efficient, in a precise, quantifiable sense. Such theorems are most straightforward to state (but not necessarily to prove, of course) in transferable-utility settings. For example, \citet{JinL2023} recently proved that in a Bayes--Nash equilibrium of a first-price auction with independent bidders' values, the expected welfare (value of the bidder who wins the item) is no less than 86\% of the expected optimal welfare (highest value).\footnote{In computer-science nomenclature, this result proves that the ``Price of Anarchy'' \citep{KoutsoupiasP1999} in this setting is at least $0.86$. The precise fraction is $1-\nicefrac{1}{e^2}$, and the same paper proves that it cannot be improved upon. This tight bound follows a sequence of papers that gradually proved tighter and tighter bounds \citep{SyrgkanisT2013,HartlineHT2014,HoyTW2018}. Inefficiency-bounding theorems indeed allow for fast-paced progress via different sets of authors building upon each other's insights through publication of gradually tighter bounds.} This inefficiency-bounding theorem allows for a more nuanced understanding of the well-known inefficiency of first-price auctions due to information asymmetries.
Other notable examples of inefficiency-bounding theorems include \citet{McAfee2002CoarseMatching,Rogerson2003SimpleMenus,Neeman2003Effectiveness,AkbarpourEtAl2023Investment}.

Inefficiency-bounding theorems can also be useful, albeit more elusive to state, in non-transferable-utility settings. For example, an early celebrated contribution of \citet{RoughgardenT2002} is that the cumulative driving time in every equilibrium of any of a large class of congestion games is no greater than $\nicefrac{4}{3}$ times the optimal cumulative driving time. This inefficiency-bounding theorem allows for further insights beyond the well-known fact \citep{Braess1968} that equilibria in many congestion games are inefficient due to unpriced externalities. We emphasize that social (in)efficiency is viewed here not as a binary (Pareto efficient vs.\ not Pareto efficient) but as a continuum. In fact, the absence of transferable utility allows some Pareto efficient alternatives to be considered less socially efficient than others (e.g., one driver driving alone on one road and all other drivers driving on the other available road can result in a very high cumulative driving time). We observe that three key properties of the cardinal measure of social inefficiency underlying this inefficiency-bounding theorem---the ratio between the cumulative driving time and the optimal cumulative driving time---contribute to its interpretability: \emph{ordinal aggregation} (over the individuals), \emph{intra-context cardinal comparability}, and \emph{inter-context cardinal comparability}.

\paragraph{Ordinal aggregation (over the individuals).} Within a congestion game, if one alternative has higher cumulative driving time than another, then the former is less socially efficient than the latter. This is driven by the disutility of every individual being exogenously cast in the same interpersonally comparable unit of measurement (minutes of travel). Ordinal aggregation might be more challenging when the (dis)utility function of each individual merely represents vNM preferences (and is thus defined only up to a monotone affine transformation). This has been the subject of research and debate within social choice theory for decades.

\vspace{-.5em}\paragraph{Intra-context cardinal comparability.} Within a specific context (specific congestion game), the social inefficiency loss between two alternatives whose cumulative driving times differ by a specific amount (e.g., five minutes) is the same as that between any other two alternatives in the same context whose cumulative driving times differ by the same amount. Intra-context cardinal comparability might be more challenging if the (dis)utility functions of the individuals are not cast in the same unit of measurement.

\vspace{-.5em}\paragraph{Inter-context cardinal comparability.} Considering two separate contexts (two congestion games), figures such as ``$\nicefrac{4}{3}$ of the optimum'' are comparably interpreted in each context. This is due to social inefficiency being not merely an aggregate utility using a common unit, but furthermore (1) measured relative to a meaningful reference point, and (2) viewed as a fraction of the optimal aggregate utility. Indeed, if one accepts that a cumulative driving time of zero is a meaningful reference point, then $\nicefrac{4}{3}$ of the optimal cumulative driving time is in a precise sense ``$\nicefrac{4}{3}$ times as bad'' (relative to the reference point) as the optimal cumulative driving time, and one might make the point that two equilibria, in two different games, that are each ``$\nicefrac{4}{3}$ times as bad'' as the optimum for its respective game are comparable in their social inefficiency. Inter-context cardinal comparability might be more challenging in the absence of a natural reference point. In such settings, even if we have a way to cardinally aggregate over the individuals using a common unit of measurement, it might not be clear how to compare across contexts (e.g., it might be hard to normalize by dividing by the optimum, since shifting any aggregate numeric function by any constant changes the ratio of function values between any alternative and the optimum).\footnote{\citet[p.~33]{LR-book} write that when restricting to ordinal comparisons, ``the non-uniqueness of the zero point is of no real concern in any of the applications of utility theory, but the arbitrary unit of measurement gives trouble.'' For inefficiency-bounding theorems, the non-uniqueness of a zero point is of real concern as well.}

\medskip

It is unclear how to state economically meaningful inefficiency-bounding theorems in settings such as matching and voting, in which a measure of social inefficiency that satisfies the three properties just discussed is not readily available.
In such settings, if we were to have at our disposal a meaningful cardinal way---that satisfies the above three properties---for measuring the social (in)efficiency of an alternative, this would open the door to using techniques, fine-tuned over several decades in the computer-science literature, to bound this (in)efficiency at alternatives of interest (such as equilibria) and derive new insights in settings in which economically interpretable inefficiency-bounding theorems might currently be lacking. 

In this paper, we initiate the study of axiomatically founding
a \emph{cardinal} social welfare function---or more specifically, 
a \emph{cardinal social inefficiency} function---satisfying the above three properties, with the aim of demonstrating a new direction for building sound economic foundations for inefficiency-bounding theorems.
Our social inefficiency function is uniquely defined (up to the unit in which it is globally measured, across all contexts) and depends only on the vNM preferences of each individual over the various possible alternatives. The definition of our social inefficiency function does not assume any exogenously given way to compare utility across individuals, nor does it assume any exogenously given reference point or disagreement point (such as having special meaning for a utility value of zero). 
We demonstrate the use of this social inefficiency function by applying it to a setting in which interpersonal comparison is notoriously challenging to justify and a reference point might not be available: Object allocation without money. In this setting, we use our social inefficiency function to leverage existing mathematical constructions and techniques from the computer-science literature to prove an approximate-efficiency result for the popular Random Serial Dictatorship mechanism.

We define a social inefficiency function as a function that for every \emph{context}---set of possible alternatives, number of individuals, and vNM preferences for the individuals over (lotteries over) these alternatives---and for every specific alternative (or lottery over alternatives), specifies a nonnegative number, called the \emph{social inefficiency} of this alternative given the context.\footnote{Considering the social inefficiency of randomized and not only deterministic alternatives allows us to quantify the social inefficiency of the outcomes of \emph{randomized} mechanisms. More generally, as \citet[p.~20]{arrow1951social} writes: ``if conceptually we imagine a choice being made between two alternatives, we cannot exclude any probability distribution over those two choices as a possible alternative.''
} After formalizing these definitions in \cref{sec:model},~in \cref{sec:axioms} we define six axioms that such a social inefficiency function should satisfy.

Our first two axioms relate to ordinal aggregation (the first of the three properties discussed above): \emph{Pareto monotonicity} is a natural desideratum on the ordinal nature of social inefficiency functions, while \emph{anonymity} is a natural desideratum on the aggregative nature of social inefficiency functions.
Our third axiom relates to intra-context cardinal comparability, and requires that the social inefficiency of lotteries be measured ex-ante, i.e., that the social inefficiency of a lottery equal the expected social inefficiency of the realized alternative.

The next two axioms relate to \emph{relative} inter-context cardinal comparability, i.e., inter-context cardinal comparability of differences between social inefficiencies. The first of these two axioms is a cardinal version of \emph{independence of irrelevant alternatives (IIA)}. The second, \emph{population-size stability}, requires that relative social inefficiencies in a context remain the same, and do not arbitrarily change, upon self-composition of multiple copies of that context; this axiom pushes our social inefficiency function to have a per-capita flavor, facilitating the comparison on equal footing of social inefficiencies in contexts with different numbers of individuals.

The final, sixth axiom relates to absolute (rather than relative) inter-context cardinal comparability: \emph{feasibility} requires that at each context, \emph{some} alternative has zero social inefficiency.

In \cref{sec:characterization}, we state the main result of the first part of this paper: That these six, logically independent axioms characterize a unique social inefficiency function up to a global choice of unit of measure, that is, up to a positive multiplicative factor applied \emph{simultaneously across all contexts and alternatives}. As we show, the characterized social inefficiency function takes a particularly natural form: It measures the per-capita additive loss in welfare relative to the maximum-welfare alternative, where welfare is defined as the sum of \emph{endogenously} normalized vNM utility functions, each utility function normalized with respect to the magnitude of the distributional conflict, that is, normalized so that its unit is the diameter of its range over the Pareto frontier.\footnote{The \emph{kindness function} of \cite{Rabin1993} also uses (albeit for a very different purpose) a normalization that depends on utility ranges over the Pareto frontier. The choice of normalization there is an explicit part of the definition rather than a consequence of an axiomatic characterization.} We therefore call this social inefficiency function the \emph{Frontier-normalized Loss in Average Welfare (FLAW)}.
FLAW has additional appealing properties beyond those explicitly required by our axioms, such as
that the social inefficiency of an alternative is not changed if dominated alternatives are added to, or removed from, the context in question.\footnote{For not satisfying (the ordinal analogue of) this property, \label{arrow-on-dominated}\citet[p.~32]{arrow1951social} criticizes the social choice function that normalizes each utility function to the range of $[0,1]$ (over all alternatives rather than over the Pareto frontier like FLAW does) and sums up the normalized utilities, calling this social choice function ``extremely unsatisfactory.''}

In \cref{sec:object-allocation}, we apply FLAW to the study of a setting in which interpersonal comparison is notoriously challenging to justify due to the absence of monetary transfers and the repugnance of measuring using money, and any normalization is challenging due to the absence of a reference point such as an outside option: Object allocation (without money). 

Random Serial Dictatorship (RSD) is a widely used allocation mechanism that is known to \emph{not} be ex-ante efficient (\citealp{Zhou1990}; see also \citealp{BogomolnaiaMoulin2001}). We apply FLAW (which measures ex-ante inefficiency) to prove that nonetheless, RSD is \emph{approximately} ex-ante efficient.
We consider the robust FLAW (i.e., worst FLAW over all object allocation problems) of truthful ordinal mechanisms at their truthtelling equilibrium (ex-ante over the randomness of the mechanism). We utilize tools that were originally developed within computer science, for the study of the ``Price of Anarchy'' \citep{FilosRatsikasFZ2014}, to prove that no such mechanism has a robust FLAW that is lower than (i.e., improves upon) that of RSD by more than~28\%. We emphasize two crucial features of FLAW that give meaning to the statement of this result: First, we can define the robust FLAW of a mechanism (recall that this is the worst FLAW over all object allocation problems) because FLAW values are cardinally comparable across contexts/object allocation problems. Second, no other mechanism being able to improve by more than 28\% is interpretable because our axioms pin down FLAW as a unique \emph{cardinal} social inefficiency function up to a \emph{global} unit of measure (and hence one mechanism's FLAW being lower by 28\% than another's is well defined independently of this unit of measure). Finally, we consider the task of computing the FLAW of any matching with respect to specific given preferences, and construct a computationally feasible algorithm for doing so.

To our knowledge, our paper is the first that, for arbitrary sets of alternatives, axiomatically microfounds a social welfare function that depends \emph{only} on the individuals' ordinal (vNM) preferences over the alternatives, whose \emph{numeric value} (rather than only its induced order) is unique (whether or not up to a global choice of unit of measure) and cannot be arbitrarily additively shifted, and which can be meaningfully compared \emph{across contexts}. We are furthermore not aware of any paper that uses or characterizes the \emph{order} induced by FLAW\@. In \cref{literature}, we survey related work. We conclude in \cref{discussion}.

There is a vast body of mathematical techniques within computer science for proving what computer scientists call \emph{approximation theorems}: theorems that ascertain that some quantity is at least a specific fraction of some reference quantity. This paper aims to make this vast body of work more applicable to economic settings by providing a microfoundation for a cardinal---rather than ordinal---aggregation of individuals' (vNM) preferences. While we focus on cardinally measuring (in)efficiency, the program presented in this paper has the potential to be applicable far more broadly. For example, one might equally imagine axiomatically microfounding a cardinal social (un)fairness function and using techniques from computer science (and beyond) to prove approximate fairness guarantees. We indeed hope that \emph{cardinal} social choice theory could facilitate bringing economic theory and theoretical computer science even closer together.

\section{Model}\label{sec:model}

A \emph{context}
is a pair $\bigl(X,(\succeq_i)_{i=1}^n\bigr)$ where $X$ is a nonempty finite
set called the set of \emph{alternatives}, $n\ge2$ is called the number of \emph{individuals}, and for each $i=1,\ldots,n$, \linebreak $\succeq_i$ is a preference order (of individual $i$) over $\Delta(X)$ that satisfies the von Neumann--Morgenstern (vNM) rationality axioms (i.e., that has a vNM expected-utility representation).\footnote{As is standard, $\Delta(X)$ denotes the set of lotteries over elements of $X$. For $x,y\in\Delta(X)$ and $0\le\alpha\le1$, we write $\alpha\cdot x+(1-\alpha)\cdot y$ to denote the lottery whose outcome is with probability $\alpha$ drawn from $x$ and otherwise drawn from $y$.}
We use the notation $\Pareto(C)\subseteq\Delta(X)$ to denote the \emph{Pareto frontier} of~$C$: the set of all $x\in\Delta(X)$ such that there does not exist $x'\in\Delta(X)$ such that $x'\succeq_i x$ for every $i$ and $x' \succ_i x$ (i.e., not $x\succeq_i x'$) for some $i$. We restrict attention to contexts in which no individual is indifferent along the entire Pareto frontier. (Such an individual does not play a real part in the distributional conflict.)\footnote{A previous version of this manuscript contained a generalized analysis that does not make this assumption.} A main example of a context is derived from an $n$-player game, where $X$ is the set of strategy profiles of the game and $\succeq_i$ is the preference order of player $i$ (over $\Delta(X)$) for every~$i$. 

\begin{definition}[Social Inefficiency Function]
A \emph{social inefficiency function} specifies for each context $C=\bigl(X,(\succeq_i)_{i=1}^n\bigr)$ and each $x\in\Delta(X)$ a number $I(C,x)\ge0$, called the \emph{social inefficiency} of $x$ with respect to $C$.\footnote{A main example is analyzing the context $C$ derived from a game and taking $x$ to be an equilibrium of the game. One might even take the supremum (or infimum) of $I(C,x)$ over all equilibria~$x$ of the game as a measure of the inefficiency of the game. This would serve a similar purpose to the Price of Anarchy from computer science, but would be invariant to strategic equivalence (i.e., not depend on the choice of vNM utility function for each player).}\textsuperscript{,}\footnote{We use the term social inefficiency \emph{function} even though, strictly speaking, $I$ is not a function because the class of all contexts is not a set. For clarity of exposition, we gloss over this point, which is immaterial for this paper because no set-theoretic issues arise here that affect our analysis.}
\end{definition}

We emphasize that the higher the social inefficiency, the \emph{less socially efficient} the alternative is viewed to be.

While the richness of the class of social preference orders induced by various social welfare functions has been repeatedly demonstrated in social choice theory, we conclude this \lcnamecref{sec:model} by demonstrating the further richness of the class of social inefficiency functions that is due to their cardinal nature.

To demonstrate this richness, we focus on \emph{relative utilitarianism} \citep{DhillonMertens1999}, which is the social preference order induced by normalizing each individual's (vNM) utility function to be between~$0$ and~$1$ over all alternatives, summing the individuals' normalized utilities to define the social welfare of a given alternative, and comparing alternatives by their social welfare.\footnote{In contrast, recall from the introduction that our social inefficiency function normalizes each utility function so that its unit is the diameter of its range over the Pareto frontier (rather than over the set of all alternatives).} Formally, using our notation, for every context $C={\bigl(X,(\succeq_i)_{i=1}^n\bigr)}$ let $u_C^1,\ldots,u_C^n$ be vNM utility representations of $\succeq_1,\ldots,\succeq_n$, respectively, such that for every $i=1,\ldots,n$, $\min_{x\in X}u^i_C(x)=0$ and $\max_{x\in X}u^i_C(x)=1$. Let $U_C(x)\eqdef\sum_{i=1}^n u_C^i(x)$. For every context $C={\bigl(X,(\succeq_i)_{i=1}^n\bigr)}$ and $x,y\in\Delta(X)$, the relative utilitarianism order is defined as $x\succeq_C y$ if and only if $U_C(x)\ge U_C(y)$. This social preference order received several axiomatic foundations (\citealp{Karni1998Impartiality,DhillonMertens1999,Segal2000Dictatorships}; see \cref{literature} for a discussion).

One social inefficiency function that induces the relative utilitarianism order is $I(C,x)=n-U_C(x)$. However, the relative utilitarianism order is also induced by many other social inefficiency functions, such as $I(C,x)\eqdef\frac{n-U_C(x)}{|X|}$ and $I(C,x)\eqdef n-U_C(x)+|X|$, and even $I(C,x)\eqdef n^n-\bigl(U_C(x)\bigr)^n$. Despite all of these social inefficiency functions being equally well motivated by all of the existing axiomatic foundations of relative utilitarianism that we are aware of (which all axiomatize only the induced order), the distinction between them is crucial to our applications (see \cref{sec:object-allocation}): $n-U_C(x)+|X|$ and $\frac{n-U_C(x)}{|X|}$ differ from $n-U_C(x)$ on inter-context cardinal comparability (e.g., when calculating the maximum of the social inefficiencies of several alternatives---each in a different context---using one of these social inefficiency functions rather than the other can change the set of worst alternatives); $n-U_C(x)+|X|$ and $n^n-\bigl(U_C(x)\bigr)^n$ differ from $n-U_C(x)$ on intra-context cardinal comparability (e.g., using one of these social inefficiency functions rather than the other can change the set of alternatives that are at least twice as inefficient as some given alternative). More generally, the relative utilitarianism order is induced by any social inefficiency function of the form $I(C,x)=f_C\bigl(-U_C(x)\bigr)$ where $f_C$ is any context-dependent monotone transformation (so long as the resulting social inefficiencies are nonnegative for all alternatives). The fact that $f_C$ can vary arbitrarily across contexts drives the richness of this class of social inefficiency functions. Our goal in this paper is to define a social inefficiency function that provides a unique cardinal comparison between alternatives, leaving open no such degrees of freedom. For more examples of social inefficiency functions (which induce different social preference orders), see \cref{sec:independence} below.

\section{Axioms}\label{sec:axioms}

In this section, we state our axioms for a social inefficiency function. The first axiom is a standard (ordinal) Pareto monotonicity axiom---i.e., if $y$ is weakly (strictly) Pareto dominated by $x$, then the social inefficiency at $y$ is weakly (strictly) greater than at $x$.

\begin{axiom}[Pareto Monotonicity]\label{axiom:paret-monotonicity}
Let $C=\bigl(X,(\succeq_i)_{i=1}^n\bigr)$ be a context. For every $x,y\in\Delta(X)$, if $x\succeq_i y$ for every $i=1,\ldots,n$, then $I(C,x)\le I(C,y)$, and if furthermore ${x\succ_i y}$ for some such $i$, then $I(C,x)<I(C,y)$.
\end{axiom}

The second axiom stipulates that the 
social inefficiency
is unchanged if the order of individuals is permuted.

\begin{axiom}[Anonymity\footnote{The following, weaker yet bulkier variant of anonymity suffices just as well for our characterization. Let $C=\bigl(X,(\succeq_i)_{i=1}^n\bigr)$ be a context, let $\pi$ be a permutation on $\{1,\ldots,n\}$, and set $C_\pi\eqdef\bigl(X,(\succeq_{\pi(i)})_{i=1}^n\bigr)$. For every $x,y\in\Delta(X)$, it holds that $I(C_\pi,x)-I(C_\pi,y)=I(C,x)-I(C,y)$. That is, the \emph{relative} social inefficiency between two alternatives is unchanged if the order of individuals is permuted.
}]\label{axiom:anonymity}
Let $C=\bigl(X,(\succeq_i)_{i=1}^n\bigr)$ be a context, let $\pi$ be a permutation on $\{1,\ldots,n\}$ and set $C_\pi\eqdef (X,(\succeq_{\pi(i)})_{i=1}^n)$. For every $x\in\Delta(X)$, it holds that $I\bigl(C_\pi,x\bigr)=I(C,x)$.
\end{axiom}

One of our main use cases for social inefficiency (and our application in this paper, in \cref{sec:object-allocation}) is quantifying the social inefficiency of the outcome of \emph{randomized} mechanisms. In such a case, it is natural to measure the ex-ante social inefficiency of the outcome. Accordingly, our third axiom stipulates that the social inefficiency of lotteries be measured ex-ante. That is, this axiom stipulates that the social inefficiency of a lottery over two alternatives is the expected social inefficiency of the realized alternative. This axiom is hence named \emph{expected inefficiency}, analogously to \emph{expected utility}.\footnote{One interpretation of this axiom (in tandem with Pareto monotonicity) is that the social inefficiency at a lottery between two alternatives is the same as at an imaginary alternative that every individual prefers as much as her certainty equivalent for the lottery. This axiom therefore allows the risk preferences \emph{of individuals} to be taken into account in the social inefficiency function, and is one of the drivers of our social inefficiency function taking a utilitarian form \citep{harsanyi1955cardinal}. While the dependence of utilitarian forms on risk preferences has drawn criticism in some settings, we refer readers to the excellent, eloquent discussion in the first five pages of the introduction of \cite{DhillonMertens1999} regarding why such criticism might be less applicable to social choice functions not intended as voting rules but rather for measurement and policy discussions. Our social inefficiency function is indeed intended for the latter, which drives our focus on cardinal comparability, both intra-context and inter-context. Another criticism of this axiom that is worth mentioning is that of \citet{Diamond1967CardinalWelfare}, who, in a nutshell, argues that it abstracts from distributional fairness considerations. We agree, but view this property as a desirable one: our social inefficiency function is indeed intended to measure the (overall) social efficiency loss rather than how the effects of this loss are distributed among the various individuals.
}

\begin{axiom}[Expected Inefficiency]\label{axiom:expected-inefficiency}
Let $C=\bigl(X,(\succeq_i)_{i=1}^n\bigr)$ be a context. For every $x,y\in\Delta(X)$ and $0\le\alpha\le1$, it holds that $I\bigl(C,\alpha\cdot x+(1-\alpha)\cdot y\bigr)=\alpha\cdot I(C,x) + {(1-\alpha)}\cdot I(C,y)$.
\end{axiom}

The fourth axiom is the familiar \emph{independence of irrelevant alternatives (IIA)}, with two adjustments. The first adjustment to IIA is required since we are not given any exogenous reference point such as a disagreement point (or worst-case alternative for society). Recall that IIA traditionally requires independence of irrelevant alternatives as long as the disagreement point is unchanged \citep{Nash1953}. In an elegant paper, \citet{Roth1977} explores variants of IIA with respect to various reference points, and in particular considers two alternative, \emph{endogenous} reference points: the \emph{ideal point}---an imaginary alternative that every individual prefers as much as her ideal alternative---and the \emph{point of minimal expectations}---an imaginary alternative that every individual prefers as much as her least-preferred \emph{undominated} alternative. Our IIA axiom requires independence of irrelevant alternatives as long as both of these points are unchanged.\footnote{We emphasize that the more points we require to be unchanged, the \emph{weaker} IIA becomes.}

The second adjustment to IIA is in what precisely is invariant to the removal of irrelevant alternatives. When the only information that an axiomatized object conveys about any two alternatives is their relative order, then IIA naturally stipulates that this order be invariant to the removal of irrelevant alternatives. In contrast, social inefficiency functions are cardinal, i.e., they are intended to convey more information than just an order. As such, the distances between the values (social inefficiencies) assigned to different alternatives must also contain salient information. (If these distances were not to contain salient information, then any monotone transformation of the social inefficiency function would convey the same information, i.e., the function would not be cardinal but rather ordinal.) Therefore, these distances should be retained when irrelevant alternatives are removed (otherwise, these alternatives are not truly irrelevant). Hence, instead of IIA requiring that if the social inefficiency of one alternative is lower than that of another then this order remains the same upon removal of irrelevant alternatives, IIA here requires that if the social inefficiency of one alternative is lower \emph{by any specific difference (distance)} than that of another then this \emph{difference} remains the same upon removal of irrelevant alternatives.

For a nonempty closed set $A\subseteq\Delta(X)$, we use $\max_{\succeq_i}A$ (resp.\ $\min_{\succeq_i}A$) to denote some $x\in A$ such that $x\succeq_i y$ (resp.\ $x\preceq_i y$) for every $y\in A$. These are well-defined since vNM preferences are continuous and since $|X|<\infty$ and hence $\Delta(X)$ is compact.

\begin{definition}[Ideal Point and Point of Minimal Expectations \citep{Roth1977}]
Let $C=\bigl(X,(\succeq_i)_{i=1}^n\bigr)$ be a context, let $\emptyset\ne X'\subset X$, and set $C'\eqdef\bigl(X',({\succeq_i}|_{\Delta(X')})_{i=1}^n\bigr)$.
\begin{itemize}
    \item $C$ and $C'$ are said to have \emph{the same ideal point} if $\max_{\succeq_i}X'\sim_i\max_{\succeq_i}X$ for every $i=1,\ldots,n$.
    \item $C$ and $C'$ are said to have \emph{the same point of minimal expectations} if \linebreak $\min_{\succeq_i}\Pareto(C')\sim_i\min_{\succeq_i}\Pareto(C)$ for every $i=1,\ldots,n$.
\end{itemize}
\end{definition}

\begin{axiom}[Independence of Irrelevant Alternatives (IIA)]\label{axiom:iia}
Let $C=\bigl(X,(\succeq_i)_{i=1}^n\bigr)$ be a context, let $\emptyset\ne X'\subset X$, and set $C'\eqdef\bigl(X',({\succeq_i}|_{\Delta(X')})_{i=1}^n\bigr)$.
If $C$ and $C'$ have the same ideal point and the same point of minimal expectations,
then for every $x,y\in\Delta(X')$ it holds that $I(C',x)-I(C',y)=I(C,x)-I(C,y)$.
\end{axiom}

Before introducing the fifth axiom, following the definition by \citet{vNM-book} of the composition of two games, we define an analogous notion of composition of several contexts into one, composed context that models the original contexts as coexisting alongside each other, with their sets of individuals made disjoint, and without any interaction between these contexts. Let $C^1={\bigl(X^1,(\succeq^1_i)_{i=1}^{n^1}\bigr)}$ and $C^2=\bigl(X^2,(\succeq^2_i)_{i=1}^{n^2}\bigr)$ be two contexts (with possibly different numbers of individuals). We define the \emph{composed context} $C^1\compose C^2$ as the context that captures (respective copies of) these two contexts existing concurrently, without affecting each other. Formally, $C^1\compose C^2\eqdef
\bigl(X^1\times X^2,(\succeq_i)_{i=1}^{n^1+n^2}\bigr)$ where for every $x,y\in
\Delta(X^1\times X^2)$, we have that:
\begin{enumerate}
    \item
For every
$i=1,\ldots,n^1$, it holds that $x\succeq_i y$ if and only if $x^1\succeq^1_i y^1$, \linebreak where $x^1,y^1\in\Delta(X^1)$ are the respective marginal distributions of $x,y$ over~$X^1$.
\item
For every
$i=n^1\!+\!1,\ldots,n^1\!+\!n^2$, it holds that $x\succeq_i y$ if and only if $x^2\succeq^2_{i-n^1}y^2$, where $x^2,y^2\in\Delta(X^2)$ are the respective marginal distributions of $x,y$ over $X^2$.
\end{enumerate}
We inductively extend this definition to define the composition of more than two contexts.

Overall, so far, \cref{axiom:paret-monotonicity,axiom:anonymity,axiom:expected-inefficiency} facilitate the comparability of the social inefficiencies of different outcomes within the same context, and \cref{axiom:iia} facilitates the comparability of social inefficiencies across contexts that have the same number of individuals but larger or smaller sets of alternatives. The fifth axiom facilitates the comparability of social inefficiencies across contexts that have different numbers of individuals, by ensuring that the relative social inefficiency between two alternatives in a context remains the same, and does not arbitrarily change, upon self-composition of (multiple copies of) the context. This axiom pushes the social inefficiency to have a per-capita flavor, but is completely silent regarding whether this is achieved by (weighted or unweighted, under many different possible norms) arithmetic averaging across individuals, geometric averaging, harmonic averaging, or in any other way.

\begin{axiom}[Population-Size Stability]
Let $C=\bigl(X,(\succeq_i)_{i=1}^n\bigr)$ be a context. For every $x,x'\in X$ and $k\in\mathbb{N}$, it holds that $I\bigl(\compose_{j=1}^k C,(x,\ldots,x)\bigr)-I\bigl(\compose_{j=1}^k C,(x',\ldots,x')\bigr)=I(C,x)-I(C,x')$.
\end{axiom}

For the sixth and final axiom, if we had, in each context, an exogenously given optimal alternative, we would have required that this alternative have a social inefficiency of zero. In the absence of such an exogenous optimal alternative, this axiom merely ensures that in each context, there is \emph{some} alternative that has a social inefficiency of zero---i.e., is considered most socially efficient.

\begin{axiom}[Feasibility]
Let $C=\bigl(X,(\succeq_i)_{i=1}^n\bigr)$ be a context. There exists $x\in\Delta(X)$ such that $I(C,x)=0$.
\end{axiom}

\section{Characterization}\label{sec:characterization}

In this section, we constructively define a social inefficiency function and prove that this social inefficiency function is uniquely characterized by our six axioms from \cref{sec:axioms}, up to a global choice of unit of measure.

\subsection{Explicit Construction}\label{sec:construction}

We start by explicitly constructing a social inefficiency function that satisfies our six axioms from \cref{sec:axioms} (see \cref{calculation} for an illustration of this construction when $n=2$). Let $C=\bigl(X,(\succeq_i)_{i=1}^n\bigr)$ be a context, and for every $i=1,\ldots,n$, let $u_i$ be some vNM utility representation of $\succeq_i$ (see \cref{calculation-context}). Let $u_i^{\mathit{max}}\eqdef\max_{x\in \Pareto(C)}u_i(x)$ and $u_i^{\mathit{min}}\eqdef\min_{x\in \Pareto(C)}u_i(x)$ (see \cref{calculation-minmax}).
Let $V(C,x)\eqdef\frac{1}{n}\sum_{i=1}^n\frac{u_i(x)-u_i^{\mathit{min}}}{u_i^{\mathit{max}}-u_i^{\mathit{min}}}$ for every $x\in\Delta(X)$ (see \cref{calculation-v}); the fractions in this expression are all well defined since no individual is indifferent along the entire Pareto frontier of $C$.
We note that $V(C,x)$ does not depend on the choice of vNM utility representations of $(\succeq_i)_{i=1}^n$. One can view $V(C,\cdot)$ as a per-capita utilitarian social welfare function that normalizes the utility function of each individual so that its maximum and minimum over the Pareto frontier of $C$ are~$1$ and~$0$, respectively. We define the social inefficiency at an alternative $x\in\Delta(X)$ as the loss in the value of $V(C,\cdot)$ compared to its maximum possible value (see \cref{calculation-i}):
\[\hat{I}(C,x)\eqdef \max_{x'\in X}V(C,x') - V(C,x).\]
\begin{figure}[p]
\tikzset{ln/.style={line width=.35pt}}
\centering
\newcommand{\axes}{%
  \draw[ln,->] (0,0) -- (5.2,0) node[below] {$u_1$};
  \draw[ln,->] (0,0) -- (0,5.2) node[left]  {$u_2$};}

\def\Pts{%
 (2.76,0.46),(1.56,0.70),(0.45,0.75),(3.87,0.86),
 (2.14,1.15),(3.20,1.48),(4.13,1.66),(1.16,1.94),
 (0.28,1.99),(2.38,2.09),(3.55,2.18),(2.79,2.65),
 (1.36,2.81),(0.35,3.22),(1.88,3.27),(2.68,3.45),
 (1.07,3.70),(1.77,3.90)}
\newcommand{\points}[1]{\foreach \p in \Pts {\fill[#1] \p circle (.9pt);} }

\def\xmin{1.77}\def\xmax{4.13}
\def\ymin{1.66}\def\ymax{3.90}

\pgfmathsetmacro{\m}{(\ymin-\ymax)/(\xmax-\xmin)}
\pgfmathsetmacro{\oneMinusM}{1-\m}

\pgfmathsetmacro{\bVzero}{\ymax - \m*\xmax}
\pgfmathsetmacro{\bVone }{\ymin - \m*\xmin}
\pgfmathsetmacro{\bVhalf}{\ymax - \m*\xmin}
\pgfmathsetmacro{\gapV}{\bVzero-\bVone}

\def\umaxx{2.68}\def\umaxy{3.45}
\pgfmathsetmacro{\bIZero}{\umaxy - \m*\umaxx}
\pgfmathsetmacro{\bIOne}{\bIZero - \gapV}

\pgfmathsetmacro{\nX}{\m/sqrt(1+\m*\m)}
\pgfmathsetmacro{\nY}{-1/sqrt(1+\m*\m)}
\pgfmathsetmacro{\nPerpX}{-\nY}
\pgfmathsetmacro{\nPerpY}{ \nX}

\pgfmathsetmacro{\sRef}{(\ymax-\ymin)/(\xmax-\xmin)}
\pgfmathsetmacro{\cRef}{\ymin - \sRef*\xmin}
\pgfmathsetmacro{\denRef}{sqrt(1+\sRef*\sRef)}

\pgfmathsetmacro{\dIx}{-\m*\gapV/(1+\m*\m)}
\pgfmathsetmacro{\dIy}{ \gapV/(1+\m*\m)}
\pgfmathsetmacro{\xI}{3.5}
\pgfmathsetmacro{\yI}{\m*\xI + \bIZero}
\pgfmathsetmacro{\midIx}{\xI-\dIx/2}
\pgfmathsetmacro{\midIy}{\yI-\dIy/2}
\pgfmathsetmacro{\distI}{(\sRef*\midIx - \midIy + \cRef)/\denRef}

\pgfmathsetmacro{\dVx}{-\dIx}
\pgfmathsetmacro{\dVy}{-\dIy}
\pgfmathsetmacro{\rhs}{-\distI*\denRef}
\pgfmathsetmacro{\constV}{\sRef*\dVx/2 - \bVzero - \dVy/2 + \cRef}
\pgfmathsetmacro{\xV}{(\rhs-\constV)/(\sRef-\m)}
\pgfmathsetmacro{\yV}{\m*\xV + \bVzero}

\newcommand{\diagline}[5]{%
  \pgfmathsetmacro{\xRaw}{#1/\oneMinusM}
  \pgfmathsetmacro{\xLab}{min(max(\xRaw,0.3),4.7)}
  \begin{scope}\clip (0,0) rectangle (5,5);
    \draw[ln,#2] (0,#1) -- (5,{#1+5*\m});
  \end{scope}
  \node[font=\scriptsize,text=#4,inner sep=0pt,outer sep=0pt]
        at ({\xLab+#5*\nX},{\xLab+#5*\nY}) {#3};
}

\begin{subfigure}{.456\linewidth}
\centering
\begin{tikzpicture}[scale=1.1]
  \useasboundingbox (0,0) rectangle (5,5);
  \axes
  \points{black}
  \draw[ln,red,dotted] (\xmin,\ymax) -- (\umaxx,\umaxy) -- (\xmax,\ymin) node [font=\scriptsize,pos=0.3,anchor=west] {$\Pareto(C)$};
\end{tikzpicture}
\caption{A two-individual context. Scale and shift of vNM utility functions are chosen arbitrarily and do not affect calculation. Pareto frontier denoted by a dotted line.}\label{calculation-context}
\end{subfigure}
\hfill
\begin{subfigure}{.456\linewidth}
\centering
\begin{tikzpicture}[scale=1.1]
  \useasboundingbox (0,0) rectangle (5,5);
  \draw[ln,red,dashed] (\xmin,0)--(\xmin,5);
  \draw[ln,red,dashed] (\xmax,0)--(\xmax,5);
  \draw[ln,red,dashed] (0,\ymin)--(5,\ymin);
  \draw[ln,red,dashed] (0,\ymax)--(5,\ymax);

  \node[font=\scriptsize,text=red,anchor=north] at (\xmax,0) {$u_1^{\max}\vphantom{u_1^{\min}}$};
  \node[font=\scriptsize,text=red,anchor=north] at (\xmin,0) {$u_1^{\min}$};
  \node[font=\scriptsize,text=red,anchor=east]  at (0,\ymax) {$u_2^{\max}$};
  \node[font=\scriptsize,text=red,anchor=east]  at (0,\ymin) {$u_2^{\min}$};

  \axes \points{black}
\end{tikzpicture}
\caption{$u_i^{\max}$ and $u_i^{\min}$ are the respective maximum and minimum utilities along the Pareto frontier, for each individual~$i$.\\}\label{calculation-minmax}
\end{subfigure}

\vspace{3em}

\begin{subfigure}{.456\linewidth}
\centering
\begin{tikzpicture}[scale=1.1]
  \useasboundingbox (0,0) rectangle (5,5);
  \draw[ln,dashed,gray] (\xmin,0)--(\xmin,5) (\xmax,0)--(\xmax,5);
  \draw[ln,dashed,gray] (0,\ymin)--(5,\ymin) (0,\ymax)--(5,\ymax);

  \node[font=\scriptsize,gray,anchor=north] at (\xmax,0) {$u_1^{\max}\vphantom{u_1^{\min}}$};
  \node[font=\scriptsize,gray,anchor=north] at (\xmin,0) {$u_1^{\min}$};
  \node[font=\scriptsize,gray,anchor=east]  at (0,\ymax) {$u_2^{\max}$};
  \node[font=\scriptsize,gray,anchor=east]  at (0,\ymin) {$u_2^{\min}$};

  \draw[ln,red,<->,>=stealth]
        (-0.1,\ymin)--(-0.1,\ymax)
        node[midway,rotate=90,above]{\tiny ~~Normalized $u_2$ unit};
  \draw[ln,red,<->,>=stealth]
        (\xmin,-0.1)--(\xmax,-0.1)
        node[midway,below,shift={(0,-.05)}]{\tiny Normalized $u_1$ unit~~~~};

  \diagline{\bVone}{red,dash pattern=on 4pt off 3pt}{$V=0$}{red}{0.4}
  \diagline{\bVhalf}{red,dash pattern=on 4pt off 3pt}{$V=\tfrac12$}{red}{0.4}
  \diagline{\bVzero}{red,dash pattern=on 4pt off 3pt}{$V=1$}{red}{0.4}

  \begin{scope}\clip (0,0) rectangle (5,5);
    \draw[ln,red,<-,>=stealth] (\xV+1.7*\nPerpX,\yV+1.7*\nPerpY) -- ++(\dVx,\dVy);
  \end{scope}
  \node[font=\scriptsize,text=red,inner sep=0pt,outer sep=0pt,
        rotate={atan2(\dVy,\dVx)+180}]
        at (\xV+\dVx/2+1.9*\nPerpX,\yV+\dVy/2+1.9*\nPerpY) {$\Delta(V)=1$};

  \axes \points{black}
\end{tikzpicture}
\caption{Each utility function is normalized by respectively normalizing $u_i^{\max}$ and $u_i^{\min}$ to $1$ and $0$. At any alternative, $V$ is the average normalized utility of the individuals.}\label{calculation-v}
\end{subfigure}
\hfill
\begin{subfigure}{.456\linewidth}
\centering
\begin{tikzpicture}[scale=1.1]
  \useasboundingbox (0,0) rectangle (5,5);
  \draw[ln,dashed,gray] (\xmin,0)--(\xmin,5) (\xmax,0)--(\xmax,5);
  \draw[ln,dashed,gray] (0,\ymin)--(5,\ymin) (0,\ymax)--(5,\ymax);

  \draw[ln,gray,<->,>=stealth]
        (-0.1,\ymin)--(-0.1,\ymax)
        node[midway,rotate=90,above]{\tiny ~~Normalized $u_2$ unit};
  \draw[ln,gray,<->,>=stealth]
        (\xmin,-0.1)--(\xmax,-0.1)
        node[midway,below,shift={(0,-.05)}]{\tiny Normalized $u_1$ unit~~~~};

  \node[font=\scriptsize,gray,anchor=north] at (\xmax,0) {$u_1^{\max}\vphantom{u_1^{\min}}$};
  \node[font=\scriptsize,gray,anchor=north] at (\xmin,0) {$u_1^{\min}$};
  \node[font=\scriptsize,gray,anchor=east]  at (0,\ymax) {$u_2^{\max}$};
  \node[font=\scriptsize,gray,anchor=east]  at (0,\ymin) {$u_2^{\min}$};

  \diagline{\bVone}{gray,dash pattern=on 4pt off 3pt}{$V=0$}{gray}{0.4}
  \diagline{\bVhalf}{gray,dash pattern=on 4pt off 3pt}{$V=\tfrac12$}{gray}{0.4}
  \diagline{\bVzero}{gray,dash pattern=on 4pt off 3pt}{$V=1$}{gray}{0.4}

  \begin{scope}\clip (0,0) rectangle (5,5);
    \draw[ln,gray,<-,>=stealth] (\xV+1.7*\nPerpX,\yV+1.7*\nPerpY) -- ++(\dVx,\dVy);
  \end{scope}
  \node[font=\scriptsize,text=gray,inner sep=0pt,outer sep=0pt,
        rotate={atan2(\dVy,\dVx)+180}]
        at (\xV+\dVx/2+1.9*\nPerpX,\yV+\dVy/2+1.9*\nPerpY) {$\Delta(V)=1$};

  \diagline{\bIZero}{red,dashdotted}{$\hat I=0$}{red}{-0.4}
  \diagline{\bIOne}{red,dashdotted}{$\hat I=1$}{red}{-0.4}

  \begin{scope}\clip (0,0) rectangle (5,5);
    \draw[ln,red,->,>=stealth] (\xI,\yI) -- ++(-\dIx,-\dIy);
  \end{scope}
  \node[font=\scriptsize,text=red,inner sep=0pt,outer sep=0pt,
        rotate={atan2(\dIy,\dIx)}]
        at (\xI-\dIx/2+0.2*\nPerpX,\yI-\dIy/2+0.2*\nPerpY) {$\Delta(\hat I)=-\Delta(V)=1$};

  \axes \points{black}
\end{tikzpicture}
\caption{The FLAW of the alternative(s) with maximum $V$ is $\hat{I}=0$. For every other alternative, its FLAW $\hat{I}$ is the loss in $V$ compared to the alternative(s) with $\hat{I}=0$.}\label{calculation-i}
\end{subfigure}

\vspace{1em}

\caption{Illustration of FLAW calculation.}\label{calculation}
\end{figure}%
We note that $\hat{I}(C,x)\ge0$ for every context $C$ and $x\in\Delta(X)$. Hence (and since $\hat{I}(C,x)$ does not depend on the choice of vNM utility representations of $(\succeq_i)_{i=1}^n$), we have that $\hat{I}$ is a social inefficiency function. 

The social inefficiency function $\hat{I}$ has a natural interpretation: It is the per-capita additive utility loss (compared to the first best) when utility functions are endogenously normalized with respect to the magnitude of the distributional conflict, that is, normalized so that the unit of each utility function is the diameter of its range over the Pareto frontier. We therefore call $\hat{I}$ the \emph{Frontier-normalized Loss in Average Welfare (FLAW)}.

It is straightforward to verify that FLAW satisfies our six axioms. An additional appealing property of FLAW is that for every $x\in\Delta(X)$, the individuals' preferences over the Pareto frontier of $C$ and over~$x$ (but not over the rest of $\Delta(X)$) uniquely determine the FLAW of~$x$---a property that we term \emph{invariance to dominated alternatives}.

\begin{definition}[Invariance to Dominated Alternatives]
A social inefficiency function~$I$ satisfies \emph{invariance to dominated alternatives} if
for every context $C={\bigl(X,(\succeq_i)_{i=1}^n\bigr)}$ and every $X'\subset X$ such that $\Delta(X')$ weakly dominates $X\setminus X'$,\footnote{That is, for every $x\in X\setminus X'$ there exists $x'\in\Delta(X')$ such that $x'\succeq_i x$ for every $i=1,\ldots,n$.} it holds that $I(C',x)=I(C,x)$ for every $x\in\Delta(X')$, where $C'\eqdef\bigl(X',({\succeq_i}|_{\Delta(X')})_{i=1}^n\bigr)$.
\end{definition}

Invariance to dominated alternatives---which we show in \cref{invariance-dominated} in \cref{properties} below to be implied by Pareto monotonicity, IIA, and feasibility---is a desirable property, which relieves models that use social inefficiency functions that satisfy it (like FLAW) from the burden of, e.g., encompassing ``universally worst'' alternatives such as ``all the members of the society die'' \citep{KanekoN1979} or having a set of alternatives that is ``sufficiently encompassing as to include, besides the actual alternatives of interest, each person's best and worst alternative within the `universal' set'' (\citealp{DhillonMertens1999}; for a discussion of these papers and other papers that make similar assumptions, see \cref{literature}). Instead, invariance to dominated alternatives allows models to focus only on modeling undominated alternatives and other alternatives of interest.\footnote{See \cref{arrow-on-dominated} (on Page~\pageref{arrow-on-dominated}) for \citeauthor{arrow1951social}'s \citeyearpar{arrow1951social} criticism of a social choice function for not satisfying invariance to dominated alternatives.}

If we had at our disposal vNM utility functions that are exogenously normalized in a way that allows for interpersonal comparison, then a social inefficiency function that measures the loss in welfare compared to the welfare-maximizing alternative might naturally depend only on the (interpersonally comparable) utility values at the welfare-maximizing alternative and at the alternative whose inefficiency is evaluated.\footnote{This is the case for the Price of Anarchy in computer science.} Invariance to dominated alternatives is a natural generalization of this property for settings in which an exogenous way to compare utilities across individuals is not available. Indeed, invariance to dominated alternatives precisely requires that the social inefficiency of $x$ is uniquely determined by the individuals' preferences over~$x$ as well as over all alternatives that are ``potential welfare maximizers,'' since the Pareto frontier is precisely the set of all alternatives that are each a welfare-maximizer for \emph{some} choice of vNM utility representations of the individuals' respective preferences (see, e.g., \citealp{mwg}, Prop.\ 16.E.2).

Another desirable property of FLAW (which is a direct consequence of Pareto monotonicity) is that zero FLAW is only possible on a (many times strict) subset of the Pareto frontier, with non-Pareto-frontier outcomes having positive FLAW\@. FLAW therefore refines the notion of Pareto efficiency.

Our main result is that our six axioms uniquely define FLAW, up to a single positive (finite) multiplicative constant that applies to all contexts and alternatives, i.e., up to a global choice of unit of measure.

\begin{theorem}\label{characterization}
A social inefficiency function $I$ satisfies Pareto monotonicity, anonymity, 
expected inefficiency, IIA, population-size stability, and feasibility if and only if there exists a constant $c>0$ such that for every context $C=\bigl(X,(\succeq_i)_{i=1}^n\bigr)$ and $x\in\Delta(X)$, it holds that $I(C,x)=c\cdot\hat{I}(C,x)$. Furthermore, dropping any of these six axioms invalidates this statement.
\end{theorem}

\subsection[Cardinal Interpretation via Nonmonetary Compensating Differentials]{Cardinal Interpretation via\texorpdfstring{\\}{ }Nonmonetary Compensating Differentials}\label{sec:interpretation}

Our social inefficiency function, FLAW ($\hat{I}$), conveys both ordinal and cardinal information. In this section, we show that the cardinal information has a natural interpretation in terms of the ordinal information via what we call \emph{nonmonetary compensating differentials}.

In what follows, for every context $C=\bigl(X,(\succeq_i)_{i=1}^n\bigr)$, we denote by $\succeq_C$ the social preference order induced by FLAW over $\Delta(X)$ (i.e., $x\succeq_C y$ iff $\hat{I}(C,x)\le\hat{I}(C,y)$), and we denote by $\mathrm{opt}_C$ an arbitrary $\succeq_C$-maximal alternative in $X$.

Recall from \cref{characterization} that our axioms characterize FLAW up to a global choice of unit of measure. Therefore, we do not attempt to interpret the (unit-dependent) value of the FLAW of any specific alternative in isolation. Rather, we interpret the unit-independent part of FLAW, which consists of proportional comparisons between FLAW values; i.e., we interpret what it means that the FLAW of one alternative is some specific fraction of the FLAW of another alternative.

We start by considering two alternatives, $x^1,x^2\in\Delta(X)$ in the same context $C=\bigl(X,(\succeq_i)_{i=1}^n\bigr)$. Assume w.l.o.g.\ that $x^1\succeq_C x^2$. An immediate natural interpretation of the FLAW of $x^1$ being an $\alpha$ fraction of the FLAW of $x^2$ (where $\alpha\in[0,1]$) is as follows. $\hat{I}(C,x^1)=\alpha\hat{I}(C,x^2)$ if and only if
\begin{equation}\label{native-interpretation}
x^1~~\sim_C~~\alpha\cdot x^2+(1-\alpha)\cdot\mathrm{opt}_C.
\end{equation}
In other words, society is indifferent between $x^1$ and a lottery of an $\alpha$ probability of $x^2$ and otherwise $\mathrm{opt}_C$. Mixing $\mathrm{opt}_C$ into each side of \cref{native-interpretation} in proportion $\alpha\!:\!1$, we obtain that $\hat{I}(C,x^1)=\alpha\hat{I}(C,x^2)$ if and only if
\begin{equation}\label{compensating-differential-intra}
\frac{1}{1+\alpha}\cdot x^1+\frac{\alpha}{1+\alpha}\cdot\mathrm{opt}_C~~\sim_C~~\frac{1}{1+\alpha}\cdot\mathrm{opt}_C+\frac{\alpha}{1+\alpha}\cdot x^2.
\end{equation}
\cref{compensating-differential-intra} captures, using only $\succeq_C$ (and $\mathrm{opt}_C$, which is itself defined using $\succeq_C$), the idea that going from $\mathrm{opt}_C$ (the first summand on the right-hand side) to $x^1$ (the first summand on the left-hand side) is ``$\alpha$~times as socially bad'' as going from $\mathrm{opt}_C$ (the second summand on the left-hand side) to $x^2$ (the second summand on the right-hand side). \cref{compensating-differential-intra} captures this by stating that these two changes, with the latter scaled by $\alpha$, socially cancel each other out. In other words, \cref{compensating-differential-intra} says that the movement from $\mathrm{opt}_C$ on the left-hand side to $x^2$ on the right-hand side, when weighted by $\alpha$, forms a (nonmonetary) compensating differential for moving from $x^1$ on the left-hand side to $\mathrm{opt}_C$ on the right-hand side. This provides a natural interpretation of $\hat{I}(C,x^1)=\alpha\hat{I}(C,x^2)$, using only the order~$\succeq_C$. We note that \cref{compensating-differential-intra} holds for any $x^1,x^2\in\Delta(X)$ and $\alpha\ge0$ such that $\hat{I}(C,x^1)=\alpha\hat{I}(C,x^2)$, without the restriction of $\alpha\in[0,1]$ that is required for \cref{native-interpretation} to hold.

The main reason we view the interpretation through \cref{compensating-differential-intra} as a more salient interpretation of the cardinal information in FLAW than the interpretation through \cref{native-interpretation} is that the former extends also to comparison across contexts, i.e., to settings in which alternative $x^1$ is in some context~$C^1$ and alternative $x^2$ is in a different context $C^2$. In this case, as we now show, there is a natural sense in which $\hat{I}(C^1,x^1)=\alpha\hat{I}(C^2,x^2)$ if and only if society is indifferent between $\frac{1}{1+\alpha}\cdot x^1+\frac{\alpha}{1+\alpha}\cdot\mathrm{opt}_{C^2}$ and $\frac{1}{1+\alpha}\cdot\mathrm{opt}_{C^1}+\frac{\alpha}{1+\alpha}\cdot x^2$ (analogously to \cref{compensating-differential-intra}). To make this precise, we must explain the meaning of lotteries that combine alternatives from both $C^1$ and $C^2$, and moreover, the sense in which the social order induced by FLAW is defined between such lotteries. We therefore start with a definition of when two contexts are naturally thought of as coexisting within a single large context.

We call two contexts with the same number of individuals $C^1={\bigl(X^1,(\succeq^1_i)_{i=1}^n\bigr)}$ and $C^2=\bigl(X^2,(\succeq^2_i)_{i=1}^n\bigr)$ \emph{compatible} if $X^1\cap X^2=\emptyset$ and there exists a context $U(C^1,C^2)={\bigl(X^1\cup X^2,(\succeq_i)_{i=1}^n\bigr)}$ such that the following two conditions hold for every $j=1,2$:
\begin{enumerate}
    \item $\succeq_i|_{\Delta(X^j)}=\succeq^j_i$ for every $i=1,\ldots,n$,
    \item $U(C^1,C^2)$ and $C^j$ have the same ideal point and the same point of minimal expectations (i.e., the two focal points of $\succeq_{U(C^1,C^2)}$ coincide with those of~$\succeq_{C^j}$). That is, both $\max_{\succeq_i}(X^1\cup X^2) \sim_i \max_{\succeq^j_i}X^j$ and $\min_{\succeq_i}\Pareto\bigl(U(C^1,C^2)\bigr) \sim_i \min_{\succeq^j_i}\Pareto(C^j)$ for every $i=1,\ldots,n$.
\end{enumerate}
Note that if such $U(C^1,C^2)$ exists, then it is uniquely defined.

The compensating-differential interpretation of the cardinal information in FLAW extends to alternatives in different compatible contexts $C^1,C^2$ through their embeddings in $U(C^1,C^2)$:
Given two compatible contexts $C^1={\bigl(X^1,(\succeq^1_i)_{i=1}^n\bigr)}$ and $C^2={\bigl(X^2,(\succeq^2_i)_{i=1}^n\bigr)}$, for every $x^1\in\Delta(X^1)$ and $x^2\in\Delta(X^2)$, we have that $\hat{I}(C^1,x^1)=\alpha\hat{I}(C^2,x^2)$ (where $\alpha\ge0$) if and only if in $U(C^1,C^2)$, going from $\mathrm{opt}_{C^1}$ to $x^1$ is ``$\alpha$ times as socially bad'' as going from $\mathrm{opt}_{C^2}$ to~$x^2$:
\[
\frac{1}{1+\alpha}\cdot x^1+\frac{\alpha}{1+\alpha}\cdot\mathrm{opt}_{C^2}~~\sim_{U(C^1,C^2)}~~\frac{1}{1+\alpha}\cdot\mathrm{opt}_{C^1}+\frac{\alpha}{1+\alpha}\cdot x^2
\]
(where each side is a lottery in $\Delta(X^1\cup X^2)$), completely analogously to \cref{compensating-differential-intra}.

When $C^1,C^2$ are incompatible contexts with the same number of individuals, an interpretation via the transitive closure of the above relation trivially applies.\footnote{That is, for contexts $C^1,C^2$ and respective alternatives $x^1,x^2$, for every sequence of contexts $C^1=D^1,D^2,\ldots,D^{k-1},D^k=C^2$ in which every two consecutive contexts are compatible, and for every sequence of alternatives $x^1=y^1,y^2,\ldots,y^{k-1},y^k=x^2$ in these respective contexts, if for every $j=1,\ldots,k-1$, going from $\mathrm{opt}_{D^j}$ to $y^j$ is ``$\alpha^j$ times as socially bad'' as going from $\mathrm{opt}_{D^{j+1}}$ to $y^{j+1}$ (i.e., $\hat{I}(D^j,y^j)=\alpha^j\hat{I}(D^{j+1},y^{j+1})$), then $\hat{I}(C^1,x^1)=\alpha^1\cdot\ldots\cdot\alpha^{k-1}\hat{I}(C^2,x^2)$.} Importantly, this transitive closure stabilizes after only two steps, connecting any two alternatives, in any two contexts with the same number of individuals, through an intermediate context, without appealing to multiplication of other ratios of inefficiencies. That is, for every two contexts (compatible or incompatible, identical or distinct) $C^1={\bigl(X^1,(\succeq^1_i)_{i=1}^n\bigr)}$ and $C^2=\bigl(X^2,(\succeq^2_i)_{i=1}^n\bigr)$ there exists a context $C^3=\bigl(X^3,(\succeq^3_i)_{i=1}^n\bigr)$ that is compatible with $C^1$ and with $C^2$ such that the following holds for every $x^1\in\Delta(X^1)$ and $x^2\in\Delta(X^2)$.
$\hat{I}(C^1,x^1)=\alpha\hat{I}(C^2,x^2)$ if and only if there exists an alternative $x^3\in\Delta(X^3)$ such that both of the following hold:\footnote{One way to construct such $C^3$ is as follows. Let $L\eqdef\max\bigl\{\hat{I}(C^j,x)~\big|~j\in\{1,2\}~\&~x\in\Delta(X^j)\bigr\}$. Assume without loss of generality that $0,1,\ldots,n\notin X^1\cup X^2$, let $X^3\eqdef\{0,1,\ldots,n\}$, and define $C^3\eqdef{\bigl(X^3,(\succeq^3_i)_{i=1}^n\bigr)}$ by defining $\succeq^3_i$ for every individual $i=1,\ldots,n$ through the vNM utility representation $u_i$ that for every $x\in X^3$ satisfies
\[
u_i(x)\eqdef
\begin{cases}
-L & x=0, \\
1 & x=i, \\
0 & \text{otherwise}.
\end{cases}
\]
The alternative $x^3$ can then be taken as the lottery over the alternatives $0\in X^3$ and $1\in X^3$ that satisfies $\hat{I}(C^3,x^3)=\hat{I}(C^2,x^2)$. Such a lottery exists since $\hat{I}(C^3,1)=0\le \hat{I}(C^2,x^2)\le L< \hat{I}(C^3,0)$. See the proof of \cref{characterization-per-size} below for an in-depth analysis of a similar construction.
}
\begin{enumerate}
\item Going from $\mathrm{opt}_{C^1}$ to $x^1$ is ``$\alpha$ times as socially bad'' as going from $\mathrm{opt}_{C^3}$ to~$x^3$:
\[
\frac{1}{1+\alpha}\cdot x^1+\frac{\alpha}{1+\alpha}\cdot\mathrm{opt}_{C^3}~~\sim_{U(C^1,C^3)}~~\frac{1}{1+\alpha}\cdot\mathrm{opt}_{C^1}+\frac{\alpha}{1+\alpha}\cdot x^3.
\]
\item Going from $\mathrm{opt}_{C^3}$ to $x^3$ is ``as socially bad'' (i.e., $\alpha=1$) as going from $\mathrm{opt}_{C^2}$ to~$x^2$:
\[
\frac{1}{2}\cdot x^3+\frac{1}{2}\cdot\mathrm{opt}_{C^2}~~\sim_{U(C^3,C^2)}~~\frac{1}{2}\cdot\mathrm{opt}_{C^3}+\frac{1}{2}\cdot x^2.
\]
\vspace{1em}
\end{enumerate}

Finally, for two contexts with different numbers of individuals $C^1={\bigl(X^1,(\succeq^1_i)_{i=1}^{n^1}\bigr)}$ and $C^2=\bigl(X^2,(\succeq^2_i)_{i=1}^{n^2}\bigr)$, the compensating-differential interpretation of the cardinal information in FLAW is again by embedding these two contexts in one large context. Because the two contexts have different numbers of individuals, such embedding requires equalizing the number of individuals in these contexts. This is done by duplicating each context an appropriate number of times so that the number of individuals in each of the two duplicated contexts is the same. That is, we let $\bar{C}^1\eqdef\compose_{j=1}^{n^2} C^1$ and $\bar{C}^2\eqdef\compose_{j=1}^{n^1} C^2$ (note that each of these contexts has $n^1n^2$ individuals). For every $x^1\in\Delta(X^1)$ and $x^2\in\Delta(X^2)$ we have that $\hat{I}(C^1,x^1)=\alpha\hat{I}(C^2,x^2)$ if and only if $\hat{I}\bigl(\bar{C}^1,(x^1,\ldots,x^1)\bigr)=\alpha\hat{I}\bigl(\bar{C}^2,(x^2,\ldots,x^2)\bigr)$,\footnote{Here, for every $j=1,2$, we abuse notation by writing $(x^j,\ldots,x^j)$ for any lottery over alternatives in $\bar{C}^j$ whose marginals all equal $x^j$.} where the latter is interpreted as above via compensating differentials for two contexts with the same number of individuals (i.e., through $U(\bar{C}^1,\bar{C}^2)$ if $\bar{C}^1$ and $\bar{C}^2$ are compatible, and otherwise through $U(\bar{C}^1,C^3)$ and then $U(C^3,\bar{C}^2)$ for a context $C^3$ that is compatible with both $\bar{C}^1$ and $\bar{C}^2$).\footnote{The same interpretation is also obtained using any other pair of duplications of $C^1$ and $C^2$ that have the same number of individuals. For example, one might define each $\bar{C}^j$ as the duplication of $C^j$ whose number of individuals is the least common multiple of $n^1$ and $n^2$.}

Overall, the social inefficiency function that emerges from our axioms, FLAW, is such that, based only on its ordinal information (its induced social order in every context), nonmonetary compensating differentials provide a complete interpretation of its unit-invariant cardinal information (proportional comparisons between FLAW values, even across contexts)---i.e., an interpretation of the cardinal comparison between alternatives that is uniquely defined by our axioms.

\subsection{Logical Independence of Our Axioms}\label{sec:independence}

In this section, we show that our six axioms are logically independent. That is, if we drop any one of them, \cref{characterization} ceases to hold (as claimed in the second part of that \lcnamecref{characterization}). We do so by demonstrating, for each of the six axioms, a social inefficiency function that satisfies all of our axioms except that axiom.

\begin{example}[Pareto Monotonicity is Logically Independent of the Other Axioms]
The constant zero social inefficiency function, $I(C,x)=0$, satisfies all of our axioms except Pareto monotonicity.
\end{example}

\begin{example}[Anonymity is Logically Independent of the Other Axioms]
Using the notation from the definition of $\hat{I}$, for every context $C$, first set $V'(C,x)\eqdef\frac{1}{1-2^{-n}}\sum_{i=1}^n2^{-i}\cdot\frac{u_i(x)-u_i^{\mathit{min}}}{u_i^{\mathit{max}}-u_i^{\mathit{min}}}$ for all $x\in\Delta(X)$ and then $I(C,x)=\max_{x'\in X}V'(C,x') - V'(C,x)$. This social inefficiency function satisfies all of our axioms except anonymity.
\end{example}

\begin{example}[Expected Inefficiency is Logically Independent of the Other Axioms]
Using the notation from the definition of $\hat{I}$, for every context $C$, first set $V'(C,x)\eqdef\frac{1}{n}\sum_{i=1}^n\bigl(\frac{u_i^{\mathit{max}}-u_i(x)}{u_i^{\mathit{max}}-u_i^{\mathit{min}}}\bigr)^2$ for all $x\in\Delta(X)$ and then $I(C,x)=V'(C,x)-\min_{x'\in\Delta(X)}V'(C,x')$. This social inefficiency function satisfies all of our axioms except expected inefficiency.
\end{example}

The remaining three (counter)examples of social inefficiency functions all induce, for each context, the same order over its alternatives as is induced by FLAW, and yet these examples differ cardinally from FLAW (and from each other). This once again showcases the richness of the design space of social inefficiency functions and that axiomatizing a cardinal social welfare function is a far stricter requirement than axiomatizing the order that it induces.

\begin{example}[IIA is Logically Independent of the Other Axioms]
The social inefficiency function $I(C,x)=\rad\bigl(|X|\bigr)\cdot\hat{I}(C,x)$ (where $X$ is the set of alternatives in $C$ and $\rad\bigl(|X|\bigr)$, the radical of $|X|$, is the product of the distinct prime factors of $|X|$) satisfies all of our axioms except IIA.
\end{example}

\begin{example}[Population-Size Stability is Logically Independent of the Other Axioms]
The social inefficiency function $I(C,x)=2^n\cdot\hat{I}(C,x)$ (where $n$ is the number of individuals in $C$) satisfies all of our axioms except population-size stability.\footnote{As part of our proof of \cref{characterization}, in \cref{anonymous-characterization-per-size} we characterize all social inefficiency functions that satisfy all of our axioms except population-size stability.}
\end{example}

\begin{example}[Feasibility is Logically Independent of the Other Axioms]
The shifted social inefficiency function $I(C,x)=1+\hat{I}(C,x)$ satisfies all of our axioms except feasibility.
\end{example}

\subsection{Proofs}

In this section, we prove \cref{characterization}, as well as a number of intermediate results, some of which might be of independent interest. Several proofs are relegated to \cref{proofs-characterization}.

\subsubsection{General Properties}\label{properties}

In this section, we state and prove several notable properties of all social inefficiency functions that satisfy various subsets of our axioms. In addition to using these properties throughout the proof of \cref{characterization} below, these properties are also of independent interest and might be viewed as additional desirable properties of social inefficiency functions that are implied by our axioms.

The first property, which immediately follows from Pareto monotonicity, is \emph{Pareto indifference}:
Two alternatives between which all individuals are indifferent must have the same social inefficiency.

\begin{definition}[Pareto Indifference]
A social inefficiency function $I$ satisfies \emph{Pareto indifference} if for every context $C=\bigl(X,(\succeq_i)_{i=1}^n\bigr)$ and every $x,y\in\Delta(X)$ such that $x\sim_i y$ for every $i=1,\ldots,n$, it holds that $I(C,x)=I(C,y)$.
\end{definition}

\begin{lemma}\label{pareto-indifference}
If a social inefficiency function satisfies Pareto monotonicity, then it satisfies Pareto indifference.
\end{lemma}

Recall from \cref{sec:construction} the definition of invariance to dominated alternatives: for every $x\in\Delta(X)$, the individuals' preferences over the Pareto frontier of $C$ and over~$x$ (but not over the rest of $\Delta(X)$) uniquely determine the social inefficiency of $x$. As noted in that \lcnamecref{sec:construction}, this property is implied by Pareto monotonicity, IIA, and feasibility.

\begin{lemma}\label{invariance-dominated}
If a social inefficiency function satisfies Pareto monotonicity, IIA, and feasibility, then it satisfies invariance to dominated alternatives.
\end{lemma}

We next establish that any social inefficiency function that satisfies Pareto indifference and invariance to dominated alternatives (or alternatively, by \cref{pareto-indifference,invariance-dominated}, any social inefficiency function that satisfies Pareto monotonicity, IIA, and feasibility) also satisfies \emph{neutrality}: it must be invariant to renaming of the alternatives in any context. To formally define neutrality, we first define what it means for two contexts to be \emph{isomorphic}, i.e., identical up to renaming of the alternatives.

\begin{definition}[Isomorphic Contexts]
Let $C^1\!=\!{\bigl(X^1,(\succeq^1_i)_{i=1}^{n}\bigr)}$ and $C^2\!=\!{\bigl(X^2,(\succeq^2_i)_{i=1}^{n}\bigr)}$ be two contexts with the same number of individuals. Let $\phi:X^1\rightarrow X^2$ be a bijection between $X^1$ and $X^2$. We extend $\phi$ in the natural way to be defined over $\Delta(X^1)$, making it a bijection between $\Delta(X^1)$ and $\Delta(X^2)$. We say that $\phi$ is an \emph{isomorphism} between $C^1$ and $C^2$ if for every $x,y\in\Delta(X^1)$ and $i=1,\ldots,n$, it holds that $x\succeq^1_i y$ if and only if $\phi(x)\succeq^2_i\phi(y)$. If there exists an isomorphism between $C^1$ and $C^2$, then we say that $C^1$ and $C^2$ are \emph{isomorphic}.
\end{definition}

\begin{definition}[Neutrality]
A social inefficiency function $I$ satisfies \emph{neutrality} if for every pair $C^1={\bigl(X^1,(\succeq^1_i)_{i=1}^{n}\bigr)}$ and $C^2={\bigl(X^2,(\succeq^2_i)_{i=1}^{n}\bigr)}$ of isomorphic contexts with isomorphism $\phi$ between $C^1$ and $C^2$ and for every $x\in\Delta(X^1)$, it holds that $I(C^1,x)=I\bigl(C^2,\phi(x)\bigr)$.
\end{definition}

\begin{lemma}\label{neutrality}
If a social inefficiency function satisfies Pareto indifference and invariance to dominated alternatives, then it satisfies neutrality.
\end{lemma}

\subsubsection{Proof of Theorem~\ref{characterization}}

To prove \cref{characterization}, we go over our axioms, one at a time, at each point characterizing the social inefficiency functions that satisfy all axioms until that point, culminating in the characterization of the social inefficiency functions that satisfy all of our axioms, i.e., \cref{characterization}.

The first step of our proof of \cref{characterization}, which is formalized in the following \lcnamecref{characterization-per-context}, has the following logic: expected inefficiency implies that a social inefficiency function is affine; adding Pareto monotonicity implies negative slopes; and adding feasibility pinpoints the affine shift so that the minimum inefficiency in the context is precisely zero. We note that while Pareto monotonicity immediately implies negative slopes when the image of $\Delta(X)$ in utility space has full dimension, the proof of this claim when this image is not of full dimension is more subtle, and employs a separating-hyperplane theorem.

\begin{lemma}\label{characterization-per-context}
For every context $C={\bigl(X,(\succeq_i)_{i=1}^n\bigr)}$, fix some vNM utility representations $u_C^1,\ldots,u_C^n$ of $\succeq_1,\ldots,\succeq_n$.
A social inefficiency function $I$ satisfies Pareto monotonicity, expected inefficiency, and feasibility if and only if for every context $C={\bigl(X,(\succeq_i)_{i=1}^n\bigr)}$ there exist constants ${c_C^1,\ldots,c_C^n>0}$ such that
\[
I(C,x)=\max_{x'\in X}\left\{\sum_{i=1}^n c_C^i u_C^i(x')\right\} - \sum_{i=1}^n c_C^i u_C^i(x)
\]
for every $x\in\Delta(X)$.
\end{lemma}

Our next step is to apply IIA\@. Assume that for two given contexts $C,C'$ with the same number of individuals and with disjoint sets of alternatives, a new context with the same number of individuals and the union of the sets of alternatives can be constructed, such that the new context and each of the original contexts have the same ideal point and the same point of minimal expectations. In this setting, for each individual $i$, if we define the slopes of $i$ in these two contexts, $-c^i_C$ and $-c^i_{C'}$ (in the notation of \cref{characterization-per-context}), with respect to utility functions that coincide on the ideal point of each original context and the point of minimal expectations of each original context, then IIA implies that these two slopes are identical. While such a new context cannot be constructed for any arbitrary pair of contexts with the same number of individuals, it can be constructed for any pair of such contexts in which one of the contexts is a specific canonical context that we define for this number of individuals. Therefore, for each individual, the slopes of the social inefficiency function in any two contexts with the same number of individuals (if they are defined with respect to such utility functions) are the same. This logic is formalized in the following \lcnamecref{characterization-per-size}, which characterizes all social inefficiency functions that satisfy the four axioms used so far: Pareto monotonicity, expected inefficiency, IIA, and feasibility.

\begin{lemma}\label{characterization-per-size}
A social inefficiency function $I$ satisfies Pareto monotonicity, expected inefficiency, IIA, and feasibility if and only if for every $n\ge2$ there exist constants $c^1_n,\ldots,c^n_n>0$ such that
\[
I(C,x)=\max_{x'\in X}\left\{\sum_{i=1}^nc^i_n\frac{u_i(x')-u_i^{\mathit{min}}}{u_i^{\mathit{max}}-u_i^{\mathit{min}}}\right\}-\sum_{i=1}^nc^i_n\frac{u_i(x)-u_i^{\mathit{min}}}{u_i^{\mathit{max}}-u_i^{\mathit{min}}}
\]
for every context with $n$ individuals $C={\bigl(X,(\succeq_i)_{i=1}^n\bigr)}$, for every choice of vNM utility representations $u_1,\ldots,u_n$ of $\succeq_1,\ldots,\succeq_n$, and for every $x\in\Delta(X)$.
\end{lemma}

Our next step is to apply anonymity. This axiom implies that for symmetric contexts---ones in which every permutation of the individuals results in an isomorphic context---the slopes of the social inefficiency function (as defined in the discussion preceding \cref{characterization-per-size}) are all the same. Hence, by \cref{characterization-per-size}, this must be the case for all contexts. This step is formalized in the following \lcnamecref{anonymous-characterization-per-size}, which characterizes all social inefficiency functions that satisfy all of our axioms except population-size stability. In particular, fixing any number of individuals $n$, this \lcnamecref{anonymous-characterization-per-size} characterizes FLAW as the social inefficiency function for contexts with $n$ individuals that uniquely (up to the unit in which it is globally measured, across all such contexts) satisfies Pareto monotonicity, anonymity, expected inefficiency, IIA, and feasibility.

\begin{lemma}\label{anonymous-characterization-per-size}
A social inefficiency function $I$ satisfies Pareto monotonicity, anonymity, expected inefficiency, IIA, and feasibility if and only if for every $n\ge2$ there exists a constant $c_n>0$ such that $I(C,x)=c_n\cdot\hat{I}(C,x)$ for every context $C={\bigl(X,(\succeq_i)_{i=1}^n\bigr)}$ with $n$ individuals and for every $x\in\Delta(X)$.
\end{lemma}

The final step is to leverage population-size stability (the final axiom that we have not used so far) to prove that the constants from \cref{anonymous-characterization-per-size} are the same for every number of individuals, characterizing $\hat{I}(C,x)$ as the unique social inefficiency function (up to the unit in which it is globally measured, across all contexts) that satisfies all six of our axioms, as claimed by \cref{characterization}. This step is formalized in the appendix.

\section{Application to Object Allocation}\label{sec:object-allocation}

In this section, we apply our social inefficiency function, FLAW, to a setting in which due to the absence of monetary transfers and outside options, and to the repugnance of measuring using money,\footnote{See \cite{Roth2007} for 
a survey of repugnance as a constraint in market design.} interpersonal comparison is challenging, and meaningful precise cardinal values (that is, not up to a monotone affine transformation) for utilities are challenging to justify: the object allocation setting. We explore this setting both from a theoretical perspective and from a computational one. Several proofs are relegated to \cref{proofs-object-allocation}.

\subsection{Model}

In an object allocation context---to which, in line with the literature on matching theory, we henceforth refer as an \emph{object allocation problem}---there are $n$ individuals and $n$ objects for some $n\ge2$. The set of alternatives $X=X(n)$~is the set of all the perfect matchings of the $n$ objects to the $n$ individuals. The individuals' vNM preferences $\succeq=(\succeq_i)_{i=1}^n$ over (lotteries over) the matchings are such that each individual's preferences are induced by vNM preferences over (lotteries over) the object allocated to this individual, with no ties between any two objects. We denote the set of all object allocation problems of size $n$ (i.e., all contexts $C={\bigl(X,(\succeq_i)_{i=1}^n\bigr)}$ such that $X=X(n)$ and the preferences are as above) by $\mathcal{M}_n$. We denote the set of all object allocation problems by $\mathcal{M}=\cup_{n=2}^{\infty}\mathcal{M}_n$.\footnote{Since we require that an object allocation problem be a context, we exclude the possibility of individuals who are indifferent along the entire Pareto frontier. Given a setting with such individuals, unless there is no distributional conflict at all (i.e., each individual has a distinct most-preferred object), if we remove these individuals---and for each of them, the only object she receives in every allocation on the Pareto frontier---we obtain an essentially equivalent object allocation problem that is in $\mathcal{M}$.}

An \emph{ordinal (matching) mechanism} is a mapping that takes a profile of arbitrary size $n$ of (strict ordinal) rankings over $n$ (pure) objects and outputs a distribution in~$\Delta\bigl(X(n)\bigr)$. Given an object allocation problem $\bigl(X,(\succeq_i)_{i=1}^n\bigr)\in\mathcal{M}$, an ordinal mechanism $\mu$ induces a game between the $n$ individuals, in which each individual~$i$ reports a ranking~$R_i$ of the $n$ objects, and the outcome is the mechanism outcome $\mu\bigl((R_i)_{i=1}^n\bigr)\in\Delta(X)$.
An ordinal mechanism is \emph{truthful} if for every object allocation problem $\bigl(X,(\succeq_i)_{i=1}^n\bigr)\in\mathcal{M}$, for each individual $i$, \emph{truthtelling} (i.e., playing the ranking induced by $\succeq_i$) is a dominant strategy in the induced game. To reduce clutter, we abuse notation by writing $\mu(\succeq)$ for the outcome of $\mu$ when every individual $i$ reports the order induced by $\succeq_i$.

\emph{Random Serial Dictatorship (RSD)} \citep{AbdulkadirogluSonmez1998} is a truthful ordinal mechanism that has found wide use in allocating ``objects'' as varied as housing units (by both governments and colleges), internships, and school slots, just to name a few examples. This mechanism, in a uniformly random order over the individuals, matches each individual to the not-yet-matched object that is ranked highest on her reported ranking.

\subsection{Robust FLAW Guarantees}

The widely used Random Serial Dictatorship (RSD) mechanism is known to \emph{not} be ex-ante efficient (\citealp{Zhou1990}; see also \citealp{BogomolnaiaMoulin2001}).\footnote{\citet{Manea2009Asymptotic} further proves that RSD is frequently not ex-ante efficient. \citet[Proposition 1]{DillenbergerSegal2025} prove that RSD is structurally far from all ex-ante efficient mechanisms.}
In this section, we apply FLAW---which measures ex-ante inefficiency---to prove that RSD is nonetheless \emph{approximately} ex-ante efficient.

We define the \emph{robust FLAW} of a truthful ordinal mechanism to be the worst FLAW, over all object allocation problems, of the truthtelling equilibrium of that mechanism (ex-ante over the randomness of the mechanism). We prove that it is impossible for any truthful ordinal mechanism to have a robust FLAW that is less than $\sim72\%$ of that of RSD.

\begin{theorem}\label{approximate-optimality}
There does not exist any ordinal mechanism $\mu$ such that
\[\sup_{(X,\succeq)\in\mathcal{M}}\hat{I}\bigl((X,\succeq),\mu(\succeq)\bigr)<
\frac{1}{2\ln 2}\sup_{(X,\succeq)\in\mathcal{M}}\hat{I}\bigl((X,\succeq),RSD(\succeq)\bigr).
\]
\end{theorem}

We note that while the statement of \cref{approximate-optimality} does not require $\mu$ to be truthful, this \lcnamecref{approximate-optimality} is most interesting when $\mu$ is truthful, as then it considers $\mu$ at equilibrium.

We recall that ratios between FLAW values, such as the ratio bounded by \cref{approximate-optimality}, were interpreted in \cref{sec:interpretation}.
More generally, we emphasize that we would not have even been able to meaningfully interpret \cref{approximate-optimality} had we not had a social inefficiency function that both allows for comparison across contexts (giving meaning to the supremum operation) and is cardinal (giving meaning to the notion of ``$\frac{1}{2\ln 2}$~of the social inefficiency'').

In the remainder of this section, we prove two bounds that together imply \cref{approximate-optimality}: A lower bound on the robust FLAW of truthful ordinal mechanisms in \cref{lowerbound}, and an upper bound on the robust FLAW of RSD in \cref{upperbound}. In both bounds, we essentially adapt proofs from \cite{FilosRatsikasFZ2014}, who bound the ``Price of Anarchy'' of RSD and other ordinal mechanisms, that is, they bound the asymptotic order of magnitude of the ratio (rather than difference) between the sum of utilities at the mechanism output and the optimal sum of utilities. As we discuss below, these bounds of \cite{FilosRatsikasFZ2014} depend on exogenously supplied cardinal utilities for specifying both an \emph{exogenously given interpersonal comparison} to define welfare, and an \emph{exogenously given ``zero''} to give meaning to fractional comparison of welfare. These two features of their analysis---taking the ratio and having exogenous normalization for the utility functions---are common to many papers in computer science; by adapting the proofs of \cite{FilosRatsikasFZ2014} with almost no conceptual change, we also demonstrate the potential of a large body of approximation theorems in computer science to be leveraged to give approximate-optimality results for FLAW, which in contrast to standard ones in computer science, depends only on the ordinal preferences and therefore assumes nothing about interpersonal comparison or an exogenously given, special zero utility value such as that of an outside option or a disagreement point.

\subsubsection{Preliminaries}

Before proving our bounds, we start with a few definitions and a \lcnamecref{efficient-iff-pareto}.

When $n$ is clear from context, we denote by $\mathcal{R}$ the set of all (strict ordinal) rankings of the $n$ (pure) objects. We call an element of $\mathcal{R}$ a \emph{ranking} and an element of $\mathcal{R}^n$ a \emph{profile of rankings}, denoting a general element of $\mathcal{R}^n$ by $R=(R_i)_{i=1}^n$. For an object allocation problem $(X,\succeq)\in\mathcal{M}_n$, for each individual $i=1,\ldots,n$ denote by $R(\succeq_i)\in\mathcal{R}$ the ranking of the objects that is induced by $\succeq_i$, and denote $R(\succeq)=\bigl(R(\succeq_i)\bigr)_{i=1}^n\in\mathcal{R}^n$.

We say that a matching $x\in X=X(n)$ is \emph{ex-post Pareto efficient} with respect to a profile of rankings $R\in\mathcal{R}^n$ if there does not exist $x'\in X$ that, according to~$R$, is ranked weakly higher than $x$ by all individuals and strictly higher than $x$ by some individual. When the object allocation problem $(X,\succeq)$ discussed is clear from context, we say that a matching is \emph{ex-post Pareto efficient} as a shorthand for saying that it is ex-post Pareto efficient with respect to $R(\succeq)$. An ordinal mechanism $\mu$ is \emph{ex-post Pareto efficient} if for every profile of rankings~$R$, it is the case that $\mu(R)$ is supported only on matchings that are ex-post Pareto efficient with respect to $R$. It is well known that RSD is ex-post Pareto efficient.

We emphasize that ex-post Pareto efficiency is defined with respect to Pareto dominance by pure matchings, whereas the Pareto frontier of $(X,\succeq)$ is defined with respect to Pareto dominance by randomized matchings. Nonetheless, for object allocation these two concepts are in fact equivalent.

\begin{lemma}\label{efficient-iff-pareto}
Let $(X,\succeq)$ be an object allocation problem and let $x\in X$. The matching $x$ is ex-post Pareto efficient if and only if $x\in\Pareto\bigl((X,\succeq)\bigr)$. Moreover, this equivalence holds even if some individuals are indifferent along the entire Pareto frontier.
\end{lemma}

\subsubsection{A Lower Bound on the Robust FLAW of \texorpdfstring{\\}{ }Truthful Ordinal Mechanisms}

We start by proving the following lower bound on the robust FLAW of every truthful ordinal mechanism.

\begin{theorem}\label{lowerbound}
For every $n\ge2$ and every ordinal mechanism $\mu$,
\[\sup_{(X,\succeq)\in\mathcal{M}_n}\hat{I}\bigl((X,\succeq),\mu(\succeq)\bigr)\ge\frac{1}{2}-\frac{1}{n}.\]
\end{theorem}

Analogously to what we noted for \cref{approximate-optimality}, while the statement of \cref{lowerbound} does not require $\mu$ to be truthful, this \lcnamecref{lowerbound} is most interesting when $\mu$ is truthful, as then it considers $\mu$ at equilibrium.

The proof of \cref{lowerbound}, which is relegated to the appendix, is an adaptation of the proof of Lemma 5 in \cite{FilosRatsikasFZ2014}, which lower-bounds the order of magnitude of the \emph{ratio} between the optimal sum of utilities and the sum of utilities at the mechanism output in ordinal mechanisms. The preference profile used in the proof of \cref{lowerbound} in fact turns out to be slightly simpler than that in their Lemma~5.

\subsubsection{An Upper Bound on the Robust FLAW of\texorpdfstring{\\}{ }Random Serial Dictatorship}

To complement \cref{lowerbound}, we prove the following \emph{upper} bound on the robust FLAW of the Random Serial Dictatorship (RSD) mechanism.

\begin{theorem}\label{upperbound}
For every $n\ge2$,
\[
\sup_{(X,\succeq)\in\mathcal{M}_n}\hat{I}\bigl((X,\succeq),RSD(\succeq)\bigr) \le \ln 2\approx0.693.
\]
\end{theorem}

Our analysis is again heavily inspired by that of \cite{FilosRatsikasFZ2014}. As already noted, there are two main differences between FLAW and their measure of social inefficiency. First (as is common in computer science), the normalization of their utility functions is exogenous rather than endogenous: Theirs are normalized so that their minimum (over all matchings) is always $0$ and their maximum is always~$1$, which exogenizes also interpersonal comparison and is affected by preferences over objects that would never be awarded by an ex-post Pareto efficient mechanism like RSD.\footnote{As discussed in \cref{sec:model}, this is one of several social inefficiency functions that all correspond to relative utilitarianism, though \citet{FilosRatsikasFZ2014} neither use the term relative utilitarianism nor appeal to any of the axiomatizations of relative utilitarianism.} Second (as is also common in computer science), they analyze (and, for RSD, upper-bound) the ratio between the optimal and attained sum of utilities rather than the difference between the two.

We start by bridging this first difference. We do so by showing that the robust (worst-case) FLAW is smaller than the worst-case loss measured by normalizing each utility function to $[0,1]$ over the entire set of alternatives.

For every individual $i=1,\ldots,n$, let $\mathcal{S}_i$ be the set of utility functions over $X$ that are each induced by a utility function over the object allocated to individual~$i$ that assigns distinct utilities to the objects. For every $i=1,\ldots,n$ and every utility function $u_i\in\mathcal{S}_i$, we denote by $R(u_i)\in\mathcal{R}$ the ranking of the objects that is induced by~$u_i$. Let $\mathcal{S}\eqdef\mathcal{S}_1\times\cdots\times\mathcal{S}_n$. For every profile $u=(u_i)_{i=1}^n\in\mathcal{S}$ of utility functions over~$X$, we denote $R(u)=\bigl(R(u_i)\bigr)_{i=1}^n\in\mathcal{R}^n$; furthermore, for every ordinal mechanism $\mu$, to reduce clutter we abuse notation by writing $\mu(u)$ instead of $\mu\bigl(R(u)\bigr)$. For every $i=1,\ldots,n$, let $\mathcal{UR}_i\subset\mathcal{S}_i$ be the set of \emph{unit range} utility functions in~$\mathcal{S}_i$---that is, functions in~$\mathcal{S}_i$ whose minimum is $0$ and whose maximum is $1$. Let $\mathcal{UR}\eqdef\mathcal{UR}_1\times\cdots\times\mathcal{UR}_n$. Finally, for every profile $u=(u_i)_{i=1}^n$ of utility functions over $X$, we extend $u$ over $\Delta(X)$ by taking expectation and define $U_u(x)\eqdef\sum_{i=1}^n u_i(x)$ and $I_u(x)\eqdef\max_{x'\in X}U_u(x')-U_u(x)$ for every $x\in\Delta(X)$.

\begin{lemma}\label{upperbound-step1}
$\sup_{(X,\succeq)\in\mathcal{M}_n}\hat{I}\bigl((X,\succeq),\mu(\succeq)\bigr) \le 
\frac{1}{n}\sup_{u\in\mathcal{UR}} I_u\bigl(\mu(u)\bigr)$ for every ex-post Pareto efficient ordinal mechanism $\mu$ and $n\ge2$.
\end{lemma}

\begin{proof}
Given a profile $u=(u_i)_{i=1}^n\in\mathcal{S}$ and $i\in\{1,\ldots,n\}$, we say that $u_i$ is \emph{Pareto-frontier-normalized} (with respect to the profile $u$) if the maximum of~$u_i$ is $1$ and the minimum of $u_i$ over the Pareto frontier is~$0$. Denote by $\mathcal{N}$ the profiles $u=(u_i)_{i=1}^n\in\mathcal{S}$ such that $u_i$ is Pareto-frontier-normalized for every $i=1,\ldots,n$. By definition, we have that $\sup_{(X,\succeq)\in\mathcal{M}_n}{\hat{I}\bigl((X,\succeq),\mu(\succeq)\bigr)}\le 
\frac{1}{n}\sup_{u\in\mathcal{N}} I_u\bigl(\mu(u)\bigr)$. It therefore suffices to show that $\sup_{u\in\mathcal{N}} I_u\bigl(\mu(u)\bigr)\le\sup_{u\in\mathcal{UR}} I_u\bigl(\mu(u)\bigr)$. Since $\mu$ is ex-post Pareto efficient, this follows by induction from the following claim, which we now prove:
For every profile $u=(u_i)_{i=1}^n\in\mathcal{S}$ such that each $u_i$ is either Pareto-frontier-normalized with respect to $u$ or in $\mathcal{UR}_i$, for every $i=1,\ldots,n$, and for every $0<\varepsilon<\frac{1}{2}$, there exists $u'_i\in\mathcal{UR}_i$ satisfying $R(u'_i)=R(u_i)$, such that $I_u(x)\le I_{(u'_i,u_{-i})}(x)+\varepsilon$ for every matching $x\in X$ that is ex-post Pareto efficient with respect to $R(u)$. (Note that it follows from \cref{efficient-iff-pareto} that the extreme objects on the Pareto frontier for each individual do not change upon the replacement of $u_i$ with $u'_i$ since $R(u'_i)=R(u_i)$, and therefore preferences that are Pareto-frontier-normalized before this replacement remain Pareto-frontier-normalized after it.)

If $u_i\in\mathcal{UR}_i$, then we can set $u'_i=u_i$ and the proof is complete; assume therefore that $u_i\notin\mathcal{UR}_i$; since $u_i$ is Pareto-frontier-normalized, we therefore have that there exists $x\in X$ such that $u_i(x)<0$.
Let $L\eqdef\min_{x\in X}u_i(x)<0$. Define $u'_i$ as follows for every $x\in X$:
\[
u'_i(x) \eqdef \begin{cases}
    \varepsilon+(1-\varepsilon)u_i(x) & u_i(x)\ge0, \\
    \varepsilon\bigl(1-u_i(x)/L\bigr) & u_i(x)<0.
\end{cases}
\]
Note that by definition, $R(u'_i)=R(u_i)$. Furthermore, since $u_i$ is Pareto-frontier-normalized and there exists $x\in X$ such that $u_i(x)<0$, we have that $u'_i\in\mathcal{UR}_i$. Note furthermore that for every matching $x\in X$ that is ex-post Pareto efficient with respect to $R(u)$, by \cref{efficient-iff-pareto} this matching is on the Pareto frontier (with respect to $u$), and hence $0\le u'_i(x)-u_i(x)\le\varepsilon$ and therefore $0\le U_{(u'_i,u_{-i})}(x)-U_u(x)\le \varepsilon$. Fix some $x_{\mathit{max}}\in\arg\max_{x\in X}U_u(x)$, and note that $x_{\mathit{max}}$ is ex-post Pareto efficient with respect to $R(u)$. For every matching $x\in X$ that is ex-post Pareto efficient with respect to $R(u)$, we have that
\[
I_{u}(x)=
U_u(x_{\mathit{max}})-U_u(x)\le
U_{(u'_i,u_{-i})}(x_{\mathit{max}})-U_{(u'_i,u_{-i})}(x)+\varepsilon\le 
I_{(u'_i,u_{-i})}(x)+\varepsilon.\qedhere
\]
\end{proof}

The second step of the proof is essentially unchanged from \citet[Lemma~3]{FilosRatsikasFZ2014}: proving that the worst-case loss is obtained when all of the mass of the utility of each individual is near the minimum ($0$) or near the maximum ($1$) of the utility function. While \cite{FilosRatsikasFZ2014} quantify the ratio, instead of difference, between the realized and optimal welfare, their proof of this step goes through, \emph{mutatis mutandis}. For every $i=1,\ldots,n$ and every $0<\varepsilon<\frac{1}{2}$, let $\mathcal{UR}_i^\varepsilon$ be the set of utility functions $u_i\in\mathcal{UR}_i$ whose range is contained in $[0,\varepsilon)\cup(1-\varepsilon,1]$. For every $0<\varepsilon<\frac{1}{2}$, let $\mathcal{UR}^\varepsilon\eqdef\mathcal{UR}_1^\varepsilon\times\cdots\times\mathcal{UR}_n^\varepsilon$.

\begin{lemma}\label{upperbound-step2}
$\sup_{u\in\mathcal{UR}} I_u\bigl(\mu(u)\bigr)=
\sup_{u\in\mathcal{UR}^\varepsilon} I_u\bigl(\mu(u)\bigr)$ for every ordinal mechanism $\mu$, $n\ge2$, and $0<\varepsilon<\frac{1}{2}$.
\end{lemma}

The last step of the proof is to quantify the right-hand side of the expression in \cref{upperbound-step2}. This is where we genuinely encounter the second difference, highlighted above, between the analysis of \cite{FilosRatsikasFZ2014} and ours: that they quantify the ratio, instead of difference, between the realized and optimal welfare. While this difference invariably means that our calculation will differ from theirs, we manage to follow a significant part of their calculation, and only diverge after the majority of the heavy lifting. Indeed, we reuse all of the RSD-specific argumentation of \citet[Lemma~4]{FilosRatsikasFZ2014} verbatim, and only diverge from their analysis in the algebraic simplification that follows.\footnote{The proof of \cref{upperbound-step3} in the appendix explicitly marks this point of divergence from the analysis of \citet{FilosRatsikasFZ2014}.}

\begin{lemma}\label{upperbound-step3}
$\lim_{\varepsilon\to0}\sup_{u\in\mathcal{UR}^\varepsilon} I_u\bigl(RSD(u)\bigr)\le(\ln 2)\cdot n$ for every $n\ge2$.
\end{lemma}

Finally, \cref{upperbound} follows from the above three \lcnamecrefs{upperbound-step1}.

\begin{proof}[Proof of \cref{upperbound}]
Combining \cref{upperbound-step1,upperbound-step2,upperbound-step3}, we have that
\[
\smashoperator[r]{\sup_{(X,\succeq)\in\mathcal{M}_n}}\hat{I}\bigl((X,\succeq),RSD(\succeq)\bigr) \le 
\frac{1}{n}\smashoperator[r]{\sup_{u\in\mathcal{UR}}}I_u\bigl(RSD(u)\bigr)=
\frac{1}{n}\lim_{\varepsilon\to0}\smashoperator[r]{\sup_{u\in\mathcal{UR}^\varepsilon}}I_u\bigl(RSD(u)\bigr)\le\ln 2.\qedhere
\]
\end{proof}

\subsection{FLAW Computation}\label{computation}

Complementing the robust guarantee of the previous section, one might wish to compute the FLAW of some matching---such as a realized mechanism output---with respect to specific given preferences. This raises computational concerns, which we discuss and resolve in this section.

To compute the FLAW of a specific allocation in an object allocation problem, letting $u_1,\ldots,u_n$ be vNM utility representations of the individuals' preferences, we need to compute (recalling the notation of the definition of $\hat{I}$ from \cref{sec:construction}): \linebreak (1)~the per-individual maximum and minimum Pareto-frontier utilities $u_i^{\mathit{max}}$ and $u_i^{\mathit{min}}$, and (2)~the linear translation $\max_{x'\in X}V(C,x')$. Once the per-individual maximum and minimum Pareto-frontier utilities are known, the linear translation can be readily computed using any of the well-known computationally feasible algorithms for the optimal assignment problem (e.g., \citealp{Kuhn1955,Munkres1957}). For the per-individual maximum and minimum Pareto-frontier utilities, by \cref{efficient-iff-pareto} we need to compute, for each individual, her most preferred and least preferred objects among those matched to her in ex-post Pareto efficient matchings. While the former is trivially her favorite object, it is at first glance unclear how one might feasibly compute the latter. Indeed, \cite{SabanS2015} show that it is NP-complete (i.e., computationally intractable) to answer, for a given individual $i$ and object $o$, the question of whether there exists an ex-post Pareto efficient matching that matches $o$ to~$i$. We nonetheless prove that it is computationally feasible to compute each individual $i$'s least preferred object among those matched to her in ex-post Pareto efficient matchings, which in turn allows us to feasibly compute the FLAW of any allocation. Our algorithm is presented in \cref{find-min-pareto-match}.

\begin{algorithm}[t]
\caption{Finds the object least preferred by individual~$\hat{i}$ among the objects matched to $\hat{i}$ in ex-post Pareto efficient matchings. 
Runs in time polynomial in $n$ (i.e., computationally feasible).
}\label{find-min-pareto-match}
\begin{algorithmic}
\Function{FindMinParetoMatch}{$n,(\succeq_i)_{i=1}^n,\hat{i}$}
\ForAll{object $o$, from $\hat{i}$'s least preferred to $\hat{i}$'s most preferred,}
\If{\Call{Test}{$n,(\succeq_i)_{i=1}^n,\hat{i},o$}}
\State\Return{$o$}
\EndIf
\EndFor
\EndFunction
\Function{Test}{$n,(\succeq_i)_{i=1}^n,\hat{i},o$}
\State $V \gets$ all individuals except $\hat{i}$ and all objects except $o$
\State $E \gets $ all individual--object pairs $(i,o')\in V^2$ s.t.\ $o' \succ_i o$
\State $G\gets(V,E)$
\State // The next line is calculable in polynomial time
\citep{HopcroftK1973}.
\State $\sigma\gets$ max-cardinality matching in the bipartite graph $G$
\If{$|\sigma|=n-1$}
\State\Return{True}
\Else
\State\Return{False}
\EndIf
\EndFunction
\end{algorithmic}
\end{algorithm}

\begin{proposition}\label{min-po}
For every object allocation problem $\bigl(X,(\succeq_i)_{i=1}^n\bigr)\in\mathcal{M}$ and individual $\hat{i}\in\{1,\ldots,n\}$, the function call \emph{\Call{FindMinParetoMatch}{$n,(\succeq_i)_{i=1}^n,\hat{i}$}} returns the object least preferred by $\hat{i}$ among the objects matched to $\hat{i}$ in ex-post Pareto efficient matchings.
\end{proposition}

Inspecting \cref{find-min-pareto-match}, it might seem at first glance that for \cref{min-po} to hold, the function call \Call{Test}{${n,(\succeq_i)_{i=1}^n,\hat{i},o}$} must check whether object $o$ is matched to $\hat{i}$ in some ex-post Pareto efficient matching, i.e., this function call must be guaranteed to return True if and only if this is the case. Such a guarantee, however, would have contradicted the above-discussed computational intractability. Rather, the function \Call{Test}{} has a somewhat weaker guarantee, which is nonetheless strong enough for \cref{min-po} to still hold. First, a one-sided guarantee holds: if $o$ is \emph{not} matched to $\hat{i}$ in any ex-post Pareto efficient matching, then this function call indeed returns False.

\begin{lemma}\label{non-po-false}
For every object allocation problem $\bigl(X,(\succeq_i)_{i=1}^n\bigr)\in\mathcal{M}$, individual~$\hat{i}$, and object $o$, if~$o$ is not matched to~$\hat{i}$ in any ex-post Pareto efficient matching, then the function call \emph{\Call{Test}{$n,(\succeq_i)_{i=1}^n,\hat{i},o$}} returns False.
\end{lemma}

Second, while the other-sided guarantee does not hold---i.e., if $o$ is matched to~$\hat{i}$ in some ex-post Pareto efficient matching, then \Call{Test}{$n,(\succeq_i)_{i=1}^n,\hat{i},o$} might return either True or False---this guarantee \emph{does} hold when $o$ is $\hat{i}$'s least preferred object among all those matched to her in ex-post Pareto efficient matchings.

\begin{lemma}\label{min-po-true}
For every object allocation problem $\bigl(X,(\succeq_i)_{i=1}^n\bigr)\in\mathcal{M}$, individual~$\hat{i}$, and object $o$, if~$o$ is the object least preferred by $\hat{i}$ among the objects matched to $\hat{i}$ in ex-post Pareto efficient matchings, then the function call \emph{\Call{Test}{$n,(\succeq_i)_{i=1}^n,\hat{i},o$}} returns True.
\end{lemma}

Since \cref{find-min-pareto-match} traverses the objects in $\hat{i}$'s reverse order of preference, these two guarantees suffice: First, as long as the \lcnamecref{find-min-pareto-match} traverses objects that are never matched to $\hat{i}$ in ex-post Pareto efficient matchings, \Call{Test}{} returns False by \cref{non-po-false}. Second, once the \lcnamecref{find-min-pareto-match} reaches an object matched to $\hat{i}$ in some ex-post Pareto efficient matching, this object is $\hat{i}$'s least preferred such object, and then \Call{Test}{} returns True by \cref{min-po-true} and the \lcnamecref{find-min-pareto-match} returns this object and concludes.

\section{Related Work}\label{literature}

To our knowledge, our paper is the first that, for arbitrary sets of alternatives, axiomatically microfounds a social welfare function that depends \emph{only} on the individuals' ordinal (vNM) preferences over the alternatives, whose \emph{numeric value} (rather than only its induced order) is unique (whether or not up to a global choice of unit of measure) and cannot be arbitrarily additively shifted, and which can be meaningfully compared \emph{across contexts}. We are furthermore not aware of any paper that uses or characterizes the \emph{order} induced by our social welfare function, FLAW.

The question of whether and how utilities of different individuals could be aggregated
in order to assess the welfare of society is long-standing in welfare economics and social choice theory, two momentous fields whose comprehensive summary is beyond the scope of this literature review. For many years, economists have been split on whether and to what extent interpersonal comparison of utility is justified. In his \citeyear{arrow1951social} book, ``Social Choice and Individual Values,'' \citeauthor{arrow1951social} takes the stand that ``interpersonal comparison of utilities has no meaning.'' Later, \cite{harsanyi1955cardinal} argues to the contrary, for utilitarianism (taking the sum of the individuals' utilities) based on the idea that the individuals' utility functions can be meaningfully compared and aggregated. He expands social welfare functions to randomized alternatives and adopts the vNM axioms \citep{vNM-book} for social preferences. He then shows that the social welfare functions that are weighted sums of the individuals' utilities (with weights that may be arbitrarily chosen) are the only ones that are Pareto monotone and satisfy an additional axiom called independence.
Following \cite{harsanyi1955cardinal}, many other papers construct social welfare functions that take as input a profile of ordinal preferences of the individuals over the given set of alternatives and yield a social ranking of the alternatives. Among these papers, several influential ones assume given utility functions that are exogenously chosen to convey some additional information beyond only the ordinal preferences \cite[e.g.,][]{demeyer1971welfare,maskin1978theorem,sen1970interpersonal,roberts1980interpersonal,hammond2005utility}.
Our paper, in contrast to all of the above, does not assume any exogenous information except for the ordinal (vNM) preferences, and yet pinpoints \emph{unique} weights on the individuals' utility functions to be used for aggregation.

A different approach is taken by \cite{KanekoN1979}, who define the \emph{Nash social welfare function}. They assume ``the existence of a distinguished alternative [\ldots]\ which represents one of the worst 
states for all individuals that we may imagine, e.g., we may imagine all the 
members of the society die,'' and evaluate social welfare by considering (the product of) the relative increases of the individuals' utilities from this distinguished alternative. The order induced by this function (which is what they axiomatize) is robust to monotone affine transformations of the utility functions that represent the individuals' preferences, while the numeric value depends on the precise choice of utility functions (and can be furthermore arbitrarily monotonically transformed without violating their axioms). \cite{piacquadio2017fairness} assumes that alternatives are allocations of commodities and introduces an inequality-concerned ``equal-preference transfer'' axiom. He then shows that if the set of alternatives consists of a bundle that is considered worst by all individuals, the social welfare functions in a class that he terms ``opportunity-equivalent utilitarianism'' are the only ones that meet this axiom in conjunction with several other axioms. In contrast to both of these papers, our construction \emph{does not} depend on a modeling choice of a distinguished worst-case alternative or bundle for all members of society.

The closest social preference order to the order induced by FLAW is \emph{relative utilitarianism}. This social preference order is obtained by normalizing each individual's (vNM) utility function to be between~$0$ and~$1$ (over all alternatives rather than over the Pareto frontier like FLAW does) and then summing the individuals' normalized utilities to define the social index according to which alternatives are compared. This social preference order received several axiomatic foundations \citep{Karni1998Impartiality,DhillonMertens1999,Segal2000Dictatorships}. Within this literature, the closest paper to ours is \cite{DhillonMertens1999}.\footnote{\citet{Segal2000Dictatorships} takes the alternatives to be different ways to distribute a divisible endowment between the individuals. In \cite{Karni1998Impartiality}, the axiomatized social preference order is over allocations that may allocate a different (randomized) alternative to each individual. \cite{DhillonMertens1999} axiomatize the relative utilitarianism social preference order over a general fixed set of alternatives.
}
As already noted by \citet{arrow1951social} (see \cref{arrow-on-dominated} on Page~\pageref{arrow-on-dominated}), the normalization to $[0,1]$ over the entire set of alternatives makes relative utilitarianism (in contrast to FLAW) sensitive to the addition or deletion of dominated alternatives. Each of the above papers motivates this normalization differently: \citet{Karni1998Impartiality} uses an impartiality axiom to provide the basis for interpersonal utility comparisons. The other two papers do not assume interpersonally comparable utilities, but rather assume an exogenous meaningful minimum point: \citet{DhillonMertens1999} assume that ``the set of alternatives is sufficiently encompassing as to include, besides the actual alternatives of interest, each person's best and worst alternative within the `universal' set''; \citet{Segal2000Dictatorships} restricts the analysis to allocation of an endowment, and allows for the entire endowment to remain unallocated. Our paper, like the latter two papers, does not assume interpersonally comparable utilities. However, since the normalization of FLAW depends only on the Pareto frontier, which is endogenously determined for each context, we do not need to assume such structure or such richness of the set of alternatives as in the latter two papers. Like all papers preceding ours that we are aware of, all three of the above papers axiomatize an order rather than a cardinally-meaningful real-valued index (i.e., any cardinal representation of the social preference order that they axiomatize can be arbitrarily monotonically transformed), and like most previous papers, this order does not allow for comparison across contexts.\footnote{Among our axioms, the only one that is violated by all social inefficiency functions that induce the relative utilitarianism order is IIA\@. Let IIA$'$ be the axiom obtained from IIA by replacing the requirement for having the same point of minimal expectations with that of having the same worst point---defined such that $C=(X,(\succeq_i)_{i=1}^n)$ and $C'=(X',({\succeq_i}|_{\Delta(X')})_{i=1}^n)$ have \emph{the same worst point} if $\min_{\succeq_i}X'\sim_i\min_{\succeq_i}X$ for every $i=1,\ldots,n$. Our axiomatic system with IIA replaced by IIA$'$ gives rise, through an analysis similar to ours, to a social inefficiency function that induces the relative utilitarianism order. Beyond being a new characterization of relative utilitarianism (distinct from that of \citealp{DhillonMertens1999}; e.g., working for any numbers of individuals and alternatives), the resulting social inefficiency function is cardinally unique (i.e., pinpointed among all the social inefficiency functions that induce the relative utilitarianism order, which are discussed in \cref{sec:model}) up to a global choice of unit of measure, allowing for intra- and inter-context comparability. (The cardinal information in the arising social inefficiency function can be interpreted as in \cref{sec:interpretation}, replacing the point of minimal expectations with the worst point in the definition of compatible contexts.)}

Another paper that axiomatically motivates a specific normalization of utility functions is that of \cite{FleurbaeyZ2021}. In this paper, an alternative is a specification of a positive (single-dimensional) consumption level for each individual ``living'' in that particular alternative; the set of alternatives is then the set of all such consumption profiles. Individuals' preferences are over their own consumption. Assuming differentiable vNM utility functions, this paper introduces axioms that justify the normalization of the utilities of all individuals so that their marginal utilities at a poverty line are the same.
Our paper, in contrast, does not assume any structure on the set of alternatives (or on the vNM utility functions).

\cite{FleurbaeyT2014} develop a social-choice theory that allows for comparison across societies, resembling our inter-context comparisons. In their paper, a \emph{social situation} consists of some population, preferences for each individual in that population over her own consumption, and a consumption profile that specifies a consumption bundle for each individual in that population. \cite{FleurbaeyT2014} seek to ordinally compare such social situations. (They are not interested in comparison of lotteries over social situations.) On a conceptual level, while their social situations technically resemble context--alternative pairs in our paper, they seek to answer the question of which of two social situations is preferred (e.g., which country is better off), while we seek to quantify, in every context, the social inefficiency of an alternative \emph{as contextualized by the context in question}. On a technical level, this translates in their setting to each individual always having preferences over the same, structured set of consumption bundles for that individual; in contrast, our sets of alternatives are completely structureless (e.g., not even a metric or topological space) and arbitrarily change between contexts (indeed, in our setting in contrast to theirs, the social inefficiency of an alternative in some context might change if new possible alternatives are added to the context). In both \cite{FleurbaeyT2014} and \cite{FleurbaeyZ2021}, and in contrast to our paper, the set of alternatives is unbounded, and therefore, since the preferences of individuals in these papers are monotone, neither an optimal or ideal point nor a Pareto frontier of alternatives (consumption profiles) exists. Furthermore, like all preceding papers of which we are aware, \cite{FleurbaeyT2014} characterize orders that satisfy various axioms rather than characterize a cardinally-meaningful real-valued index.

A different stream of literature follows the Nash bargaining solution \citep{Nash1953} and, in a setting where society consists of two individuals, searches for a socially preferred \emph{alternative} rather than a social preference order. In this literature, it is usually assumed that such an alternative should be unique. \cite{Roth1977} considers different ``Independence of Irrelevant Alternatives (IIA)'' axioms and the bargaining solutions they obtain. While Nash's bargaining solution is obtained for the axiom of ``IIA other than the disagreement point,'' \cite{Roth1977} shows that the bargaining solution of \cite{kalai1975other} (see also \citealp{Raiffa1953}) is obtained for the axiom of ``IIA other than both the disagreement point and the ideal point.'' \cite{Roth1977} also introduces the (endogenous) point of minimal expectations, and characterizes the bargaining solutions that meet the axiom of ``IIA other than the point of minimal expectations'' or the axiom of ``IIA other than the ideal point.'' However, he does not consider, as we do, the axiom of ``IIA other than both the point of minimal expectations and the ideal point.'' \cite{nalebuff2021perspective} looks for ``perspective invariant'' bargaining solutions in the sense of obeying a modified version of Nash's axioms with respect to the disagreement point as well as a flipped version of the same axioms with respect to the ideal point.
He normalizes so that the (exogenous) disagreement point is $(0,0)$ and the (endogenous) ideal point is $(1,1)$ and exhibits such a modification of Nash's axioms that
gives a bargaining solution that maximizes the sum of the two individuals' utilities. 

Computer scientists have been interested in measuring the inefficiency of a given alternative compared to the optimal efficiency. Within games, \cite{KoutsoupiasP1999} define the Price of Anarchy as the ratio between the worst-case performance at any equilibrium and the optimal performance at any strategy profile---both measured using some welfare (or cost) function (such as the cumulative driving time of all individuals, as studied by \citealp{RoughgardenT2002}). Following that paper, the
Price of Anarchy (many times using the sum of exogenously normalized utilities as the welfare function) has been
adopted not only in computer science but also in operations research, industrial engineering, and other areas as a measure of the inefficiency of an interactive environment. For example, the Price of Anarchy has been applied to classical operations research areas such as supply chains (e.g., \citealp{perakis2007price}), queuing (see \citealp{ghosh2021inefficiency} for a survey), load balancing (e.g., \citealp{ayesta2010price}), and scheduling (e.g., \citealp{hoeksma2019price}).
In contrast to the Price of Anarchy, FLAW does not require exogenously given welfare or cost functions (or exogenously normalized utility functions) that are cardinally comparable within and between contexts.

\section{Discussion}\label{discussion}

In this paper, we axiomatically define a \emph{cardinal} social welfare function, or more specifically a cardinal social inefficiency function, which provides a unique cardinal comparison between alternatives, even across contexts. Like any axiomatization, our axioms can of course be debated. Changing one of them might lead to a very different social inefficiency function. We do not claim that our set of axioms is necessarily the only possible one or even the best possible one. Our aim is to demonstrate the applicability of the axiomatic approach for microfounding
an economically meaningful quantitative comparison of the social inefficiency of different alternatives within and across contexts.

Beyond demonstrating the usefulness of our social inefficiency function, FLAW, and the types of theorems that it can facilitate, an additional purpose of our application section, \cref{sec:object-allocation}, is to demonstrate the relative ease with which existing computer-science analyses can be adapted to yield FLAW results. For this reason, we make no attempt to develop further techniques to get a better fraction than $\frac{1}{2\ln 2}$ in \cref{approximate-optimality}. As is common in computer science, approximation theorems allow for progress to be made gradually,
and we hence
leave the question of tighter bounds in the object allocation application as an interesting open problem.\footnote{Of course, ours is not the first economics paper that derives an approximation result and leaves potentially tightening that result as an open problem. 
For example (for approximately optimizing a monopolist's revenue rather than an aggregate, social welfare function), \cite{HartN2017} prove that for a two-item monopolist, optimally pricing each item separately guarantees at least $50\%$ of the optimal revenue, a guarantee that was later improved to $\sim62\%$ by \cite{HartR2019}, with a tight bound on this guarantee still remaining an open problem.}

While we demonstrate the usefulness of FLAW within noncooperative game theory, we emphasize that its applicability is potentially much broader. Indeed, it can be applied to any setting that features a set of alternatives and preferences of individuals over them. 
For example, within a bargaining setting, FLAW essentially frames negotiations as, instead of bargaining with respect to an exogenous disagreement point \emph{{\`a} la} Nash, bargaining with respect to an \emph{endogenous} agreement \emph{region} (the Pareto frontier). This might be seen as in line with many negotiation texts for executives that discuss the ``Zone of Possible Agreement (ZOPA)'' as a key object that frames negotiations.\footnote{See, e.g., \url{https://online.hbs.edu/blog/post/understanding-zopa}. We thank Yuval Gonczarowski for introducing us to this notion.}

As noted in the introduction, our analysis in this paper is a demonstration of a potentially larger program of providing a microfoundation for what computer scientists call \emph{approximation theorems}. These theorems, which ascertain that some quantity is at least a specific fraction of some reference quantity, are not inherently computational. Nonetheless, they are emblematic of theoretical computer science, and as such, tools and techniques for deriving them have been developed over many decades. While we focus on cardinally measuring (in)efficiency, one might imagine axiomatically defining cardinal measures of other objectives of interest such as (un)fairness, (in)equity, and so forth. Such cardinal measures, of (in)efficiency or otherwise, can also be used beyond bounding the loss in a certain objective in specific alternatives, e.g., for deriving \emph{cardinal} comparative statics: Quantifying not only the direction, but also the magnitude, of change in a certain objective as certain model parameters are changed.
Overall, our paper demonstrates a new direction for providing a sound economic foundation for approximation theorems and other quantitative theorems as we enter the second quarter-century of intense EconCS cross-pollination.

\bibliographystyle{abbrvnat}
\bibliography{bib}

\clearpage

\appendix
\setcounter{page}{1} \renewcommand{\thepage}{A.\arabic{page}}

\section{Proofs Omitted from Section~\ref{sec:characterization}}\label{proofs-characterization}

\begin{proof}[Proof of \cref{pareto-indifference}]
Let $C={\bigl(X,(\succeq_i)_{i=1}^n\bigr)}$ be a context and let $x,y\in\Delta(X)$ such that $x\sim_i y$ for every $i=1,\ldots,n$. By Pareto monotonicity (applied twice) we have both that $I(C,x)\le I(C,y)$ and that $I(C,x)\ge I(C,y)$, and hence $I(C,x)=I(C,y)$, and so $I$ satisfies Pareto indifference.
\end{proof}

\begin{proof}[Proof of \cref{invariance-dominated}]
Let $C={\bigl(X,(\succeq_i)_{i=1}^n\bigr)}$ be a context and let $X'\subset X$ be a set of alternatives such that $\Delta(X')$ weakly dominates $X\setminus X'$. Let $C'\eqdef\bigl(X',({\succeq_i}|_{\Delta(X')})_{i=1}^n\bigr)$.

We first claim that there exists $\hat{x}\in\Delta(X')$ such that $I(C,\hat{x})=0$. To see this, first note that by feasibility there exists $x\in\Delta(X)$ (not necessarily in $\Delta(X')$) such that $I(C,x)=0$. Write $x=\sum_{\ell=1}^L\alpha_\ell\cdot x_\ell$ where $\alpha_1,\ldots,\alpha_L\in[0,1]$ and $x_1,\ldots,x_L\in X$. By assumption, for every $\ell=1,\ldots,L$ there exist $\hat{x}_\ell\in\Delta(X')$ such that $\hat{x}_\ell\succeq_i x_\ell$ for every $i=1,\ldots,n$. Therefore, $\hat{x}\eqdef\sum_{\ell=1}^L\alpha_\ell\cdot\hat{x}_\ell\in\Delta(X')$ satisfies $\hat{x}\succeq_i x$ for every $i=1,\ldots,n$. By Pareto monotonicity, $I(C,\hat{x})=0$ as claimed.

By feasibility, there exists $\hat{x}'\in\Delta(X')$ such that $I(C',\hat{x}')=0$. 
By construction of $C'$ and since $\Delta(X')$ weakly dominates $X\setminus X'$, the contexts 
$C$ and $C'$ have the same ideal point and the same point of minimal expectations. Hence,
by IIA we have that
\[
0\le I(C,\hat{x}')=I(C,\hat{x}')-I(C,\hat{x})=I(C',\hat{x}')-I(C',\hat{x})=-I(C',\hat{x})\le0,
\]
and therefore $I\bigl(C',\hat{x}\bigr)=0$. Now, for every $x\in\Delta(X')$, by IIA we have that
\[
I(C',x)=I(C',x)-I(C',\hat{x})=I(C,x)-I(C,\hat{x})=I(C,x).\qedhere
\]
\end{proof}

\begin{proof}[Proof of \cref{neutrality}]
Let $C^1={\bigl(X^1,(\succeq^1_i)_{i=1}^{n}\bigr)}$ and $C^2={\bigl(X^2,(\succeq^2_i)_{i=1}^{n}\bigr)}$ be two isomorphic contexts and let $\phi$ be an isomorphism between $C^1$ and $C^2$.

We first prove that for the special case in which $X^1\cap X^2=\emptyset$, it indeed holds that $I(C^1,x)=I\bigl(C^2,\phi(x)\bigr)$ for every $x\in\Delta(X^1)$. Let $X\eqdef X^1\cup X^2$ and define a context $C=\bigl(X,(\succeq_i)_{i=1}^{n}\bigr)$ by defining $\succeq_i$ for every $i$ using the isomorphism~$\phi$ as follows: Let $x,y\in\Delta(X)$. By definition of $X$, there exist $\alpha\in[0,1]$, $x^1\in\Delta(X^1)$, and $x^2\in\Delta(X^2)$ such that $x=\alpha\cdot x^1 + (1-\alpha)\cdot x^2$. Similarly, there exist $\beta\in[0,1]$, $y^1\in\Delta(X^1)$, and $y^2\in\Delta(X^2)$ such that $y=\beta\cdot y^1 + (1-\beta)\cdot y^2$. We define $\succeq_i$ so that $x\succeq_i y$ if and only if $\alpha\cdot \phi(x^1) + (1-\alpha)\cdot x^2 \succeq^2_i \beta\cdot \phi(y^1) + (1-\beta)\cdot y^2$. (Equivalently, if and only if $\alpha\cdot x^1 + (1-\alpha)\cdot \phi^{-1}(x^2) \succeq^1_i \beta\cdot y^1 + (1-\beta)\cdot \phi^{-1}(y^2)$.) Note that $C^1={\bigl(X^1,({\succeq_i}|_{\Delta(X^1)})_{i=1}^n\bigr)}$ and that $C^2={\bigl(X^2,({\succeq_i}|_{\Delta(X^2)})_{i=1}^n\bigr)}$.

Note that for every $x\in\Delta(X^1)$, we have that $x\sim_i\phi(x)$ for every $i=1,\ldots,n$. 
By Pareto indifference,
we therefore have that $I(C,x)=I\bigl(C,\phi(x)\bigr)$. Therefore, by invariance to dominated alternatives
(applied twice), we have that
\[
I(C^1,x)=I(C,x)=I\bigl(C,\phi(x)\bigr)=I\bigl(C^2,\phi(x)\bigr),
\]
completing the proof for the special case in which $X^1\cap X^2=\emptyset$.

For the general case in which possibly $X^1\cap X^2\ne\emptyset$, let $X^3$ be a set such that $|X^3|=|X^1|$ (and hence $|X^3|=|X^2|$) and $(X^1\cup X^2)\cap X^3=\emptyset$. Since $|X^3|=|X^1|$, there exists a bijection $\psi$ between $X^3$ and $X^1$. We extend $\psi$ in the natural way to be defined over $\Delta(X^3)$, making it a bijection between $\Delta(X^3)$ and $\Delta(X^1)$. For every $i=1,\ldots,n$, we define $\succeq^3_i$ over $\Delta(X^3)$ using the bijection~$\psi$ as follows: Let $x,y\in\Delta(X^3)$. We define $\succeq^3_i$ so that $x\succeq^3_i y$ if and only if $\psi(x)\succeq^1_i\psi(y)$. Note that $\psi$ is an isomorphism between $C^3\eqdef{\bigl(X^3,(\succeq^3_i)_{i=1}^{n}\bigr)}$ and $C^1$, and therefore $\phi\circ\psi$ is an isomorphism between $C^3$ and $C^2$. For every $x\in\Delta(X^1)$, we therefore have by the above special case (applied twice) that $I(C^1,x)=I\bigl(C^3,\psi^{-1}(x)\bigr)=I\bigl(C^2,\phi(x)\bigr)$, and so $I$ satisfies neutrality.
\end{proof}

\begin{proof}[Proof of \cref{characterization-per-context}]
The ``if'' direction is straightforward to check.
We prove the ``only if'' direction.

Let $C={\bigl(X,(\succeq_i)_{i=1}^n\bigr)}$ be a context and let $u_C^1,\ldots,u_C^n$ be vNM utility representations of $\succeq_1,\ldots,\succeq_n$. Denote $u_C=(u_C^1,\ldots,u_C^n)$ and let $U\eqdef u_C\bigl(\Delta(X)\bigr)$. By Pareto indifference (\cref{pareto-indifference}),
$I(C,x)$ depends on $x$ only through the values $u_C^1(x),\ldots,u_C^n(x)$, i.e., there exists a function $f:U\rightarrow\mathbb{R}_{\ge0}$ such that $I(C,x)=f\bigl(u_C^1(x),\ldots,u_C^n(x)\bigr)$. By expected inefficiency and since all the functions $u_C^i$ are affine, $f$ preserves convex combinations on $U$; hence, $f$ is an affine function on $U$. Therefore, $f$ extends to an affine function on the affine hull of $U$, which in turn extends to an affine function on $\mathbb{R}^n$. Therefore, $f$ can be written in the form
\begin{equation}\label{inefficiency-affine}
f(u)=a-c\cdot u
\end{equation}
for some constants $a\in\mathbb{R}$ and $c=(c^1,\ldots,c^n)\in\mathbb{R}^n$. We emphasize that since $U$ is not necessarily of full dimension, the choice of $a,c$ need not be unique, and we fix one such choice arbitrarily. By Pareto monotonicity, $f$ is strictly decreasing, in the sense that $f(u)>f(u')$ for every $u,u'\in U$ such that $u\le u'$ and $u\ne u'$.

Let $D\eqdef\mathrm{Span}(U-U)\subseteq\mathbb{R}^n$ be the subspace spanned by all directions between utilities in $U$. Since the set $U-U$ is convex (by convexity of $U$) and symmetric, contains $0$, and spans $D$, it is absorbing in $D$. We claim that there exists $\varepsilon>0$ such that for every $d\in D\setminus\{0\}$ such that $d\ge0$, it holds that $c\cdot d\ge\varepsilon\mathbf{1}\cdot d$ (where $\mathbf{1}\eqdef(1,\ldots,1)\in\mathbb{R}^n$). Let $V\eqdef\{d\in D \mid d\ge0~\&~\mathbf{1}\cdot d=1\}$. If $V=\emptyset$, then one may choose any $\varepsilon>0$ and the claim vacuously holds. Assume, therefore, that $V\ne\emptyset$. For every $d\in V$, because $d\in D$ and since $U-U$ is absorbing in $D$, there exist $u,u'\in U$ such that $u-u'=\delta d$ for some $\delta>0$. Since $d\ne0$, we have that $u\ne u'$. Since $d\ge0$, we furthermore have that $u\ge u'$. Therefore, since $f$ is strictly decreasing, $\delta c\cdot d=c\cdot (u-u')=f(u')-f(u)>0$ and hence $c\cdot d>0$. Since $c\cdot d>0$ for every $d\in V$ and since $d\mapsto c\cdot d$ is a continuous function, by compactness of $V$ we have that there exists $\varepsilon>0$ such that $c\cdot d\ge\varepsilon$ for every $d\in V$. Since $\mathbf{1}\cdot d=1$ for every $d\in V$, we therefore have that $c\cdot d\ge\varepsilon\mathbf{1}\cdot d$ for every $d\in V$, and since this inequality is homogeneous, we have that it holds for every $d\in D\setminus\{0\}$ such that $d\ge0$ (regardless of whether or not $\mathbf{1}\cdot d=1$), as claimed.

Let $A$ be a matrix whose columns span $D^{\perp}$. We claim that there exists a vector $\lambda$ such that $A\lambda\ge\varepsilon\mathbf{1}-c$. Assume for contradiction that this is not the case. Then, by Farkas's lemma, there exists $d\ge0$ such that both $A^{\top}d=0$ and $(\varepsilon\mathbf{1}-c)\cdot d>0$. By the latter, $d\ne0$. Since $A^{\top}d=0$, by definition of $A$ we have that $d\in (D^{\perp})^{\perp}=D$. Therefore, since $d\ge0$, by the previous argument we have that $c\cdot d\ge\varepsilon\mathbf{1}\cdot d$, contradicting the conclusion of Farkas's lemma that $(\varepsilon\mathbf{1}-c)\cdot d>0$. Therefore, there indeed exists a vector $\lambda$ such that $A\lambda\ge\varepsilon\mathbf{1}-c$, as claimed.

Let $c_C\eqdef c+A\lambda$. By definition of $\lambda$, we have that $c_C\ge\varepsilon\mathbf{1}>0$. By definition of $A$, we have that $A\lambda\in D^{\perp}$. Therefore, $(A\lambda)\cdot u$ is constant for all $u\in U$. Set $a_C\eqdef a + (A\lambda)\cdot u$ for some (equivalently, every) $u\in U$. Therefore, for every $u\in U$, we have that:
\[
f(u)=a-c\cdot u=a + (A\lambda)\cdot u - (c+A\lambda)\cdot u = a_C - c_C\cdot u.
\]
Therefore, $I(C,x)=a_C - c_C\cdot u_C(x)$ for every $x\in\Delta(X)$.

Since by definition of a social inefficiency function, $I(C,x)\ge0$ for every $x\in X$, we have that $a_C\ge\max_{x'\in X}\bigl\{c_C\cdot u_C(x')\bigr\}$. By feasibility, there exists $x\in\Delta(X)$ such that $a_C-c_C\cdot u_C(x)=0$; hence, there exists a pure alternative $x'$ in the support of $x$ such that $a_C-c_C\cdot u_C(x')\le 0$. Therefore, $a_C\le\max_{x'\in X}\bigl\{c_C\cdot u_C(x')\bigr\}$. By both of these, $a_C=\max_{x'\in X}\bigl\{c_C\cdot u_C(x')\bigr\}$.
\end{proof}

\begin{proof}[Proof of \cref{characterization-per-size}]
The ``if'' direction is straightforward to check.
We prove the ``only if'' direction.

Let $n\ge2$.
Let $\hat{X}_n\eqdef\{0,1,\ldots,n\}$, and define a context $\hat{C}_n\eqdef{\bigl(\hat{X}_n,(\hat{\succeq}_i)_{i=1}^n\bigr)}$ by defining $\hat{\succeq}_i$ for every individual $i=1,\ldots,n$ through the vNM utility representation $\hat{u}_i$ that for every $x\in \hat{X}_n$ satisfies
\[
\hat{u}_i(x)\eqdef
\begin{cases}
1 & x=i, \\
0 & x\ne i.
\end{cases}
\]
(In the discussion preceding \cref{characterization-per-size}, this context is referred to as the ``canonical context with $n$ individuals.'')
By \cref{characterization-per-context}, there exist constants $c^1_{\hat{C}_n},\ldots,c^n_{\hat{C}_n}>0$ such that $I(\hat{C}_n,x)=\max_{x'\in\hat{X}_n}\bigl\{\sum_{i=1}^n c^i_{\hat{C}_n} \hat{u}_i(x')\bigr\} - \sum_{i=1}^n c^i_{\hat{C}_n} \hat{u}_i(x)$ for every $x\in\Delta(\hat{X}_n)$. Denote $c^i_n\eqdef c^i_{\hat{C}_n}$ for every $i=1,\ldots,n$.

Let $C={\bigl(X,(\succeq_i)_{i=1}^n\bigr)}$ be a context with $n$ individuals. By neutrality (\cref{neutrality}, where Pareto indifference is by \cref{pareto-indifference} and invariance to dominated alternatives is by \cref{invariance-dominated}),
assume without loss of generality that $X\cap\hat{X}_n=\emptyset$.

Let $X'\eqdef X\cup\hat{X}_n$ and define a context $C'\eqdef\bigl(X',(\succeq'_i)_{i=1}^{n}\bigr)$ where for every~$i=1,\ldots,n$, the preference $\succeq'_i$ is obtained from $\succeq_i$ by stipulating for every $x\in \hat{X}_n$ that $x\sim'_i\max_{\succeq_i}X$ if $\hat{u}_i(x)=1$ and $x\sim'_i\min_{\succeq_i}\Pareto(C)$ if $\hat{u}_i(x)=0$. Note that $\hat{C}_n=\bigl(\hat{X}_n,({\succeq'_i}|_{\Delta(\hat{X}_n)})_{i=1}^n\bigr)$ as well as that $\hat{C}_n$ and $C'$ have the same ideal point and the same point of minimal expectations.

For every $i=1,\ldots,n$, choose a vNM utility representation $u'_i$ of $\succeq'_i$ that coincides with $\hat{u}_i$ over~$\hat{X}_n$.
By \cref{characterization-per-context}, there exist constants $c^1_{C'},\ldots,c^n_{C'}>0$ such that $I(C',x)=\max_{x'\in X'}\bigl\{\sum_{i=1}^n c^i_{C'}u'_i(x')\bigr\} - \sum_{i=1}^n c^i_{C'}u'_i(x)$ for every $x\in\Delta(X')$. For every $i=1,\ldots,n$, 
by IIA we have that \[
c^i_{C'}=I(C',0)-I(C',i)=I(\hat{C}_n,0)-I(\hat{C}_n,i)=c^i_{\hat{C}_n}=c^i_n.
\]

For every $i=1,\ldots,n$, let $u_i$ be some vNM utility representation of $\succeq_i$. Observing that $\Delta(X)$ weakly dominates $\hat{X}_n$ in $C'$, by invariance to dominated alternatives (\cref{invariance-dominated}),
for every $x\in\Delta(X)$ we have that
\begin{multline*}
I(C,x)=I(C',x)=
\max_{x'\in X}\left\{\sum_{i=1}^n c^i_{C'}u'_i(x')\right\} - \sum_{i=1}^n c^i_{C'}u'_i(x)=\\
\max_{x'\in X}\left\{\sum_{i=1}^n c^i_n u'_i(x')\right\} -
\sum_{i=1}^n c^i_n u'_i(x)= \\
\max_{x'\in X}\left\{\sum_{i=1}^n c^i_n\cdot\frac{u_i(x')-u_i^{\mathit{min}}}{u_i^{\mathit{max}}-u_i^{\mathit{min}}}\right\} - \sum_{i=1}^n c^i_n\cdot\frac{u_i(x)-u_i^{\mathit{min}}}{u_i^{\mathit{max}}-u_i^{\mathit{min}}}.\tag*{\qedhere}
\end{multline*}
\end{proof}

\begin{proof}[Proof of \cref{anonymous-characterization-per-size}]
The ``if'' direction follows by construction of $\hat{I}$. We prove the ``only if'' direction.

Let $n\ge2$. For every $i=1,\ldots,n$, let $c^i_n>0$ be the constant guaranteed to exist by \cref{characterization-per-size}. Let $c_n\eqdef n\cdot c^1_n$. Let $\hat{C}_n=\bigl(\hat{X}_n,(\hat{\succeq}_i)_{i=1}^n\bigr)$ be as in the proof of \cref{characterization-per-size}.

Let $i\in\{2,\ldots,n\}$. Let $\pi$ be a permutation on $\{1,\ldots,n\}$ such that $\pi(1)=i$ and let $C'\eqdef \bigl(\hat{X}_n,(\hat{\succeq}_{\pi(i)})_{i=1}^n\bigr)$. By \cref{characterization-per-size} (applied twice) and by anonymity,
\[
c^i_n=
I(\hat{C}_n,0)-I(\hat{C}_n,i)=
I(C',0)-I(C',i)=
c^1_n=\frac{c_n}{n}.
\]

Let $C={\bigl(X,(\succeq_i)_{i=1}^n\bigr)}$ be a context with $n$ individuals. For every $i=1,\ldots,n$, let $u_i$ be some vNM utility representation of $\succeq_i$. Since we have shown that $c^i_n=\frac{c_n}{n}$ for all $i=1,\ldots,n$, by \cref{characterization-per-size} we have for every $x\in\Delta(X)$ that
\[
I(C,x)=\max_{x'\in X}\left\{\sum_{i=1}^n\frac{c_n}{n}\frac{u_i(x')-u_i^{\mathit{min}}}{u_i^{\mathit{max}}-u_i^{\mathit{min}}}\right\}-\sum_{i=1}^n\frac{c_n}{n}\frac{u_i(x)-u_i^{\mathit{min}}}{u_i^{\mathit{max}}-u_i^{\mathit{min}}}=c_n\cdot\hat{I}(C,x).\qedhere
\]
\end{proof}

\begin{proof}[Proof of \cref{characterization}]
The ``if'' direction follows by construction of $\hat{I}$.
We prove the ``only if'' direction. (For the ``furthermore'' part, see \cref{sec:independence}.) For every $n\ge2$, let $c_n>0$ be the constant guaranteed to exist by \cref{anonymous-characterization-per-size}. Let $c\eqdef c_2$. By \cref{anonymous-characterization-per-size}, it suffices to prove for all integers $n\ge3$ that $c_n=c$.

Let $n\ge3$, and let $\hat{C}_n$ and $\hat{C}_2$ be as in the proof of \cref{characterization-per-size}. By definition of $\hat{I}$, by \cref{anonymous-characterization-per-size} (applied four times), and by population-size stability (applied twice),
\begin{multline*}
c_n=
c_n\cdot n\bigl(\hat{I}(\hat{C}_n,0)-\hat{I}(\hat{C}_n,1)\bigr)=
n\bigl(I(\hat{C}_n,0)-I(\hat{C}_n,1)\bigr)=\\
n\left(I\bigl(\hat{C}_n\compose\hat{C}_n,(0,0)\bigr)-I\bigl(\hat{C}_n\compose\hat{C}_n,(1,1)\bigr)\right)=\\
c_{2n}\cdot n\left(\hat{I}\bigl(\hat{C}_n\compose\hat{C}_n,(0,0)\bigr)-\hat{I}\bigl(\hat{C}_n\compose\hat{C}_n,(1,1)\bigr)\right)=
c_{2n}=\\
c_{2n}\cdot2\left(\hat{I}\bigl(\compose_{j=1}^n\hat{C}_2,(0,\ldots,0)\bigr)-\hat{I}\bigl(\compose_{j=1}^n\hat{C}_2,(1,\ldots,1)\bigr)\right)=\\
2\left(I\bigl(\compose_{j=1}^n\hat{C}_2,(0,\ldots,0)\bigr)-I\bigl(\compose_{j=1}^n\hat{C}_2,(1,\ldots,1)\bigr)\right)=\\
2\bigl(I(\hat{C}_2,0)-I(\hat{C}_2,1)\bigr)=
c_2\cdot2\bigl(\hat{I}(\hat{C}_2,0)-\hat{I}(\hat{C}_2,1)\bigr)=
c_2=
c.\tag*{\qedhere}
\end{multline*}
\end{proof}

\section{Proofs Omitted from Section~\ref{sec:object-allocation}}\label{proofs-object-allocation}

\begin{proof}[Proof of \cref{efficient-iff-pareto}]
The ``if'' direction is immediate: An alternative on the Pareto frontier is not strictly dominated by any alternative in $\Delta(X)$, and therefore also not strictly dominated by any alternative in $X$, and hence it is ex-post Pareto efficient.

For the ``only if'' direction, we need to prove that for every ex-post Pareto efficient matching $x\in X$ and for every $y\in\Delta(X)$ such that $y\succeq_i x$ for all $i=1,\ldots,n$, it holds that $y=x$. We prove this by induction over the number of individuals $n$. The case of $n=1$ is immediate. Assume that $n\ge 2$ and that the claim holds for $n-1$.

Since $x$ is ex-post Pareto efficient, there exists $\hat{i}\in\{1,\ldots,n\}$ that is matched to her favorite object in $x$ (otherwise, there would exist a Pareto-improving trading cycle, contradicting the ex-post Pareto efficiency of $x$). Since $y\succeq_{\hat{i}} x$ and since the individuals' preferences over the objects are strict, $\hat{i}$ is matched with probability $1$ to her favorite object in $y$ as well. Let $N'\eqdef\{1,\ldots,n\}\setminus\bigl\{\hat{i}\bigr\}$ and let $O'$ be the set of all objects except for the one matched to $\hat{i}$ by $x$. Let $X'$ be the set of all perfect matchings between $N'$ and $O'$, and denote by $x'\in X'$ and $y'\in\Delta(X')$ the respective restrictions of $x$ and $y$ to $N'$. Note that $x'$ is ex-post Pareto efficient with respect to $R\bigl((\succeq_i|_{\Delta(X')})_{i\in N'}\bigr)$ and that $y'\succeq_i x'$ for every $i\in N'$. By the induction hypothesis, $y'=x'$. Therefore, $y=x$.
\end{proof}

\begin{proof}[Proof of \cref{lowerbound}]
We say that $\mu$ is \emph{anonymous} if for every profile of rankings $R=(R_i)_{i=1}^n\in\mathcal{R}^n$ and for every permutation $\pi$ on $\{1,\ldots,n\}$, it holds that $\pi^{-1}\bigl(\mu(\pi(R))\bigr)=\mu(R)$, where $\pi(R)\eqdef(R_{\pi(1)},\ldots,R_{\pi(n)})$ and where $\pi^{-1}\bigl(\mu(\pi(R))\bigr)$ is defined so that if $\mu\bigl(\pi(R)\bigr)=\sum_{\ell=1}^L p_\ell\cdot x_\ell$ where $x_\ell\in X$ for all $1\le\ell\le L$, then $\pi^{-1}\bigl(\mu(\pi(R))\bigr)\eqdef\sum_{\ell=1}^L p_\ell\cdot \pi^{-1}(x_\ell)$, where $\pi^{-1}(x_\ell)$ matches $i$ to whichever object $\pi^{-1}(i)$ is matched to in $x_\ell$, for every $i=1,\ldots,n$.

We start with the case in which $\mu$ is anonymous. Let $0<\varepsilon<\nicefrac{1}{n}$. For each $i=1,\ldots,n$, define a utility function $u_i$ over objects as follows for every object $j=1,\ldots,n$:
\[
u_i(j)\eqdef\begin{cases}
1-(j-1)\varepsilon & j\le i, \\
\frac{n-j}{n}\varepsilon & j>i,
\end{cases}
\]
and let $\succeq_i$ be the preference over $\Delta(X)$ whose vNM utility representation is $u_i$. Note that $R(\succeq_i)=R(\succeq_j)$ for all $i,j\in\{1,\ldots,n\}$. Therefore, all matchings are ex-post Pareto efficient, and by \cref{efficient-iff-pareto} are on the Pareto frontier, and so $u_i^{\max}=1$ and $u_i^{\min}=0$ for every $i=1,\ldots,n\!-\!1$. Additionally, $u_n^{\max}=1$ and $u_n^{\min}=1-(n-1)\varepsilon$. Let $\hat{u}_i(\cdot)\eqdef\frac{u_i(\cdot)-u_i^{\min}}{u_i^{\max}-u_i^{\min}}$ be the Pareto-frontier-normalized utility function for every $i=1,\ldots,n$. We have that $\hat{u}_i=u_i$ for every $i=1,\ldots,n\!-\!1$, and that
\[
\hat{u}_n(j)=\frac{1-(j-1)\varepsilon-\bigl(1-(n-1)\varepsilon\bigr)}{1-\bigl(1-(n-1)\varepsilon\bigr)}=\frac{(n-j)\varepsilon}{(n-1)\varepsilon}=\frac{n-j}{n-1}
\]
for every $j=1,\ldots,n$.

Denote $\succeq\eqdef(\succeq_1,\ldots,\succeq_n)$. Since $\mu$ is ordinal and anonymous and since $R(\succeq_i)=R(\succeq_j)$ for all $i,j\in\{1,\ldots,n\}$, when the reported profile of rankings is $R(\succeq)$, every matching in $X$ is output by $\mu$ with the same probability. Therefore, the sum of Pareto-frontier-normalized utilities at $\mu(\succeq)$ is
\[
\frac{1}{n}\sum_{i=1}^n\sum_{j=1}^n\hat{u}_i(j) =
\frac{1}{n}\sum_{i=1}^{n-1}\sum_{j=1}^n\hat{u}_i(j) + \frac{1}{n}\sum_{j=1}^n\hat{u}_n(j)\le
\frac{1}{n}\sum_{i=1}^{n-1} \bigl(i + (n-i)\varepsilon\bigr) + \frac{1}{2} =
\frac{n}{2} + \frac{n-1}{2}\varepsilon.
\]

In contrast, the sum of Pareto-frontier-normalized utilities at the matching that matches each individual $i=1,\ldots,n$ to the same-indexed object $j=i$ is
\[
\sum_{i=1}^n\hat{u}_i(i) =
\sum_{i=1}^{n-1} \bigl(1 - (i-1)\varepsilon\bigr) + 0 \ge
n-1 - \frac{(n-1)(n-2)}{2}\varepsilon.
\]

Therefore, we have that\footnote{Recall that $\hat{I}$ is a per-capita measure, i.e., it averages over the individuals' utility losses.}
\begin{multline*}
\hat{I}\bigl((X,\succeq),\mu(\succeq)\bigr)\ge
\frac{1}{n}\left(n-1 - \frac{(n-1)(n-2)}{2}\varepsilon-\left(\frac{n}{2}+\frac{n-1}{2}\varepsilon\right)\right)=\\
\frac{1}{2}-\frac{1}{n}-\frac{(n-1)^2}{2n}\varepsilon\xrightarrow[\varepsilon \to 0]{}\frac{1}{2}-\frac{1}{n}.
\end{multline*}

We move on to the case in which $\mu$ is not anonymous. Let $0<\varepsilon<\nicefrac{1}{n}$. Let~$\hat{\mu}$ be such that for every $R\in\mathcal{R}^n$, it holds that
$\hat{\mu}(R)=\sum_{\pi}\frac{1}{n!}\cdot{\pi^{-1}\bigl(\mu(\pi(R))\bigr)}$, where the sum is over all permutations on $\{1,\ldots,n\}$.
By the anonymous case, there exists an object allocation problem $\bigl(X,\succeq)\in\mathcal{M}_n$ such that $\hat{I}\bigl((X,\succeq),\hat{\mu}(\succeq)\bigr)\ge{\frac{1}{2}-\frac{1}{n}-\frac{(n-1)^2}{2n}\varepsilon}$. Therefore, there exists a permutation~$\pi$ on~$\{1,\ldots,n\}$ such that ${\hat{I}\bigl((X,\succeq),\pi^{-1}(\mu(\pi(\succeq)))\bigr)}\ge{\frac{1}{2}-\frac{1}{n}-\frac{(n-1)^2}{2n}\varepsilon}$, where $\pi(\succeq)$ denotes the preference profile in which each individual $i$'s preferences over objects are those of individual $\pi(i)$ in $\succeq$. We conclude the proof by focusing on the preference profile $\pi(\succeq)$ and noting that \[\hat{I}\bigl((X,\pi(\succeq)),\mu(\pi(\succeq))\bigr)={\hat{I}\bigl((X,\succeq),\pi^{-1}(\mu(\pi(\succeq)))\bigr)}\ge\frac{1}{2}-\frac{1}{n}-\frac{(n-1)^2}{2n}\varepsilon\xrightarrow[\varepsilon \to 0]{}\frac{1}{2}-\frac{1}{n}.\]
\end{proof}

\begin{proof}[Proof of \cref{upperbound-step2}]
The result follows by induction from the following claim, which we now prove: For every profile $u=(u_i)_{i=1}^n\in\mathcal{UR}$ and $i=1,\ldots,n$ such that\footnote{We denote $u_i(X)\eqdef\bigl\{u_i(x)~\big|~x\in X\bigr\}$.} $u_i(X)\cap[\varepsilon,1-\varepsilon]\ne\emptyset$, there exists $u_i'\in\mathcal{UR}_i$ satisfying $R(u'_i)=R(u_i)$ and $\bigl|u'_i(X)\cap[\varepsilon,1-\varepsilon]\bigr|<\bigl|u_i(X)\cap[\varepsilon,1-\varepsilon]\bigr|$, such that $I_u\bigl(\mu(u)\bigr)\le I_{(u'_i,u_{-i})}\bigl(\mu(u)\bigr)$.

Let $[0,\bar{l}]\subseteq[0,\varepsilon)$, $[l,r]\subseteq[\varepsilon,1-\varepsilon]$,
and $[\bar{r},1]\subseteq(1-\varepsilon,1]$ be the smallest such segments such that $u_i(X)\subseteq[0,\bar{l}]\cup[l,r]\cup[\bar{r},1]$. These are well defined because the three intersections $u_i(X)\cap[0,\varepsilon)$, $u_i(X)\cap[\varepsilon,1-\varepsilon]$, and $u_i(X)\cap(1-\varepsilon,1]$ are all nonempty by assumption. Let $\hat{l}=\frac{\bar{l}+\varepsilon}{2}$ and $\hat{r}=\frac{\bar{r}+1-\varepsilon}{2}$. For every $\delta\in\{\hat{l}-l,0,\hat{r}-r\}$, define the utility function $u^\delta_i$ as follows for every $x\in X$:
\[
u_i^\delta(x)\eqdef\begin{cases}
    u_i(x) & u_i(x)\notin[l,r], \\
    u_i(x)+\delta & u_i(x)\in[l,r]. 
\end{cases}
\]
Note that $u_i^0=u_i$ and that $R(u_i^\delta)=R(u_i)$ for every $\delta$. Note also that for every $\delta\in\{\hat{l}-l,\hat{r}-r\}$, we have that $\bigl|u_i^\delta(X)\cap[\varepsilon,1-\varepsilon]\bigr|<\bigl|u_i(X)\cap[\varepsilon,1-\varepsilon]\bigr|$. It remains to show that $I_{(u_i^0,u_{-i)}}\bigl(\mu(u)\bigr)\le\max_{\delta\in\{\hat{l}-l,\hat{r}-r\}}\bigl\{I_{(u^\delta_i,u_{-i})}\bigl(\mu(u)\bigr)\bigr\}$.

Let $x_{\mathit{max}}\in\arg\max_{x\in X}U_u(x)$. Note that the function $f(\delta)\eqdef U_{(u_i^{\delta},u_{-i})}(x_{\mathit{max}})-U_{(u_i^{\delta},u_{-i})}\bigl(\mu(u)\bigr)$ is a linear function. Therefore, at one of the extreme values $\hat{\delta}\in\{\hat{l}-l,\hat{r}-r\}$, the function $f$ attains a weakly higher value than at the intermediate value $\delta=0$. Let $u'_i\eqdef u_i^{\hat{\delta}}$. We conclude by observing that
$I_u\bigl(\mu(u)\bigr)=f(0)\le f(\hat{\delta}) \le I_{(u'_i,u_{-i})}\bigl(\mu(u)\bigr)$.
\end{proof}

\begin{proof}[Proof of \cref{upperbound-step3}]
Let $0<\varepsilon<\nicefrac{1}{n^3}$ and let $u=(u_i)_{i=1}^n\in\mathcal{UR}^\varepsilon$. 
It is enough to prove that $I_u\bigl(RSD(u)\bigr)-2n\varepsilon<(\ln 2)\cdot n$. By definition of $\mathcal{UR}^\varepsilon$, there exists $k\in\{1,\ldots,n\}$ such that $\bigl|k-\max_{x\in X}U_u(x)\bigr|<\varepsilon n<\nicefrac{1}{n^2}$. \cite{FilosRatsikasFZ2014} prove that under these conditions, we have (in our notation) that
\[
U_u\bigl(RSD(u)\bigr)\ge\sum_{l=0}^{\lfloor k/2\rfloor}\frac{k-2l}{n-l}-n\varepsilon.
\]
Since $\max_{x\in X}U_u(x)-n\varepsilon< k$,
we have that\footnote{At this point, our analysis diverges from that of \cite{FilosRatsikasFZ2014}, who go on to bound the order of magnitude of the ratio $\frac{U_u(RSD(u))}{\max_{x\in X}U_u(x)}$.}
\[
I_u\bigl(RSD(u)\bigr)-2n\varepsilon <
k-\sum_{l=0}^{\lfloor k/2\rfloor}\frac{k-2l}{n-l}.
\]
Note that $d\bigl(\frac{k-2x}{n-x}\bigr)/dx=\frac{k-2n}{(n-x)^2}<0$. Therefore,
\[
\sum_{l=0}^{\lfloor k/2\rfloor}\frac{k-2l}{n-l}>
\int_{0}^{k/2}\frac{k-2x}{n-x}dx.
\]
Combining these two, we have that
\begin{multline*}
I_u\bigl(RSD(u)\bigr)-2n\varepsilon <
k-\int_{0}^{k/2}\frac{k-2x}{n-x}dx
=
\int_{0}^{k/2}\left(2-\frac{k-2x}{n-x}\right)dx=
\\
\int_{0}^{k/2}\frac{2n-2x-k+2x}{n-x}dx=
\int_{0}^{k/2}\frac{2n-k}{n-x}dx=
(2n-k)\int_{0}^{k/2}\frac{1}{n-x}dx=
\\
(2n-k)\Bigl(\ln n - \ln\bigl(n-\tfrac{k}{2}\bigr)\Bigr) = 
(2n-k)\left(\ln \frac{1}{1-\frac{k}{2n}}\right).
\end{multline*}
Writing $k'=\frac{k}{n}\le1$, we then have that
\[
I_u\bigl(RSD(u)\bigr)-2n\varepsilon <
n\cdot (2-k')\left(\ln\frac{1}{1-\frac{k'}{2}}\right)\le n\cdot\ln 2.\qedhere
\]
\end{proof}

\begin{proof}[Proof of \cref{non-po-false}]
Assume for contradiction that True is returned. Then, the max-cardinality matching $\sigma$ in $G$ is of size $n-1$. Therefore, $x\eqdef\sigma\cup(\hat{i},o)\in X$ is a perfect matching of the $n$ individuals to the $n$ objects such that each individual except $\hat{i}$ is matched to an object that she strictly prefers to $o$. Let $x'\in X$ be an ex-post Pareto efficient matching such that $x'\succeq_i x$ for every $i=1,\ldots,n$. Then, in $x'$, each individual except $\hat{i}$ is matched to an object that she strictly prefers to $o$. Therefore, $o$ is matched to $\hat{i}$ in $x'$, in contradiction to $o$ not being matched to $\hat{i}$ in any ex-post Pareto efficient matching.
\end{proof}

\begin{proof}[Proof of \cref{min-po-true}]
Let $x\in X$ be an ex-post Pareto efficient matching that matches $o$ to $\hat{i}$. Assume for contradiction that in $x$, some individual $j\ne\hat{i}$ is matched to an object that she strictly prefers less than $o$. Let $y\in X$ be the matching obtained from $x$ by exchanging the objects of $\hat{i}$ and $j$. Then, $j$ strictly prefers $y$ to $x$. Since $x$ is ex-post Pareto efficient, we therefore have that $\hat{i}$ is matched in $y$ to an object she strictly prefers less than~$o$. Let $y'\in X$ be an ex-post Pareto efficient matching such that $y'\succeq_i y$ for every $i=1,\ldots,n$. We claim that $\hat{i}$ is matched in $y'$ to an object she strictly prefers less than $o$; indeed, if this were not the case, then it would hold that $y'\succeq_i x$ for every $i=1,\ldots,n$ and $y'\succ_j x$, in contradiction to $x$ being ex-post Pareto efficient. We conclude that $y'$ is an ex-post Pareto efficient matching that matches to $\hat{i}$ an object that she strictly prefers less than $o$, contradicting the definition of $o$. Therefore, in $x$ each individual $j\ne\hat{i}$ is matched to an object that she strictly prefers to $o$. Hence, during the computation of the function call \Call{Test}{$n,(\succeq_i)_{i=1}^n,\hat{i},o$}, the max-cardinality matching $\sigma$ in $G$ is of size $n\!-\!1$. Thus, True is returned.
\end{proof}

\end{document}